\documentclass[11pt, reqno]{amsart}
\usepackage{amsmath,amsthm,amssymb}

\theoremstyle{plain}
\newtheorem{thm}{Theorem}[section]
\newtheorem{prop}[thm]{Proposition}
\newtheorem{lem}[thm]{Lemma}
\newtheorem{cor}[thm]{Corollary}

\theoremstyle{definition}

\newtheorem{rem}[thm]{Remark}
\newtheorem{defn}[thm]{Definition}
\newtheorem{eg}[thm]{Example}
\newtheorem{subtitle}[thm]{}
\newtheorem{ex}{Exercise}[section]
\numberwithin{equation}{section}

\def\a{\alpha}
\def\b{\beta}
\def\d{\delta}
\def\D{\triangle}

\def\g{\gamma}
\def\G{\Gamma}

\def\l{\lambda}
\def\L{\Lambda}
\def\n{\vert\/}

\def\cv{{\mathcal{V}}}

\def\li{\langle}
\def\ri{\rangle}
\def\n{|\/ }
\def\tr{{\rm tr}}
\def\bs{\bigskip}
\def\ms{\medskip}
\def\ss{\smallskip}

\def\ti{\tilde}
\def\p{\partial}

\def\I{{\rm I\/}}

\def\diag{{\rm diag}}
\def\ad{{\rm ad}}

\def\C{\mathbb{C}}

\def\R{\mathbb{R} }

\newcommand{\beq}{\begin{equation}}
\newcommand{\eeq}{\end{equation}}
\newcommand{\beg}{\begin{eg}}
\newcommand{\eeg}{\end{eg}}
\newcommand{\bthm}{\begin{thm}}
\newcommand{\ethm}{\end{thm}}
\newcommand{\bprop}{\begin{prop}}
\newcommand{\eprop}{\end{prop}}
\newcommand{\bcor}{\begin{cor}}
\newcommand{\ecor}{\end{cor}}
\newcommand{\blem}{\begin{lem}}
\newcommand{\elem}{\end{lem}}
\newcommand{\bca}{\begin{cases}}
\newcommand{\eca}{\end{cases}}
\newcommand{\brem}{\begin{rem}}
\newcommand{\erem}{\end{rem}}
\newcommand{\bpm}{\begin{pmatrix}}
\newcommand{\epm}{\end{pmatrix}}
\newcommand{\bbm}{\begin{bmatrix}}
\newcommand{\ebm}{\end{bmatrix}}
\newcommand{\bvm}{\begin{vmatrix}}
\newcommand{\evm}{\end{vmatrix}}
\newcommand{\bdefn}{\begin{defn}}
\newcommand{\edefn}{\end{defn}}
\newcommand{\bsub}{\begin{subtitle}}
\newcommand{\esub}{\end{subtitle}}
\newcommand{\bex}{\begin{ex}}
\newcommand{\eex}{\end{ex}}
\newcommand{\ben}{\begin{enumerate}}
\newcommand{\een}{\end{enumerate}}

\def\ti{\tilde}
\def\p{\partial}

\def\I{{\rm I\/}}

\def\diag{{\rm diag}}
\def\ad{{\rm ad}}

\def\bc{{\bf c}}

\def\rd{{\rm d\,}}

\def\R{\mathbb{R} }
\def\C{\mathbb{C}}

\def\calB{{\mathcal B}}

\def\calG{{\mathcal G}}
\def\calJ{{\mathcal J}}

\def\calL{{\mathcal L}}
\def\calM{{\mathcal M}}
\def\calN{{\mathcal N}}
\def\calO{{\mathcal O}}

\def\calS{{\mathcal S}}
\def\calT{{\mathcal T}}

\def\calV{{\mathcal V}}
\def\calX{{\mathcal X}}

\def\li{\langle}
\def\ri{\rangle}
\def\bu{$\bullet$}

\def\frakP{{\mathfrak{P}}}
\def\frakS{{\mathfrak{S}}}
\def\frakJ{{\mathfrak{J}}}

\def\Ker{{\rm Ker\/}}
\def\mod{{\rm mod\,}}

\begin{document}

\title[Tau-Virasoro nxn KdV]
{Tau function and Virasoro action for the $n\times n$ KdV hierarchy}
\author{Chuu-Lian Terng$^\dag$}\thanks{$^\dag$Research supported
in  part by NSF Grant DMS-1109342}
\address{Department of Mathematics\\
University of California at Irvine, Irvine, CA 92697-3875.  Email: cterng@math.uci.edu}
\author{Karen Uhlenbeck$^*$}\thanks{$^*$Research supported in part by the Sid Richardson
Regents' Chair Funds, University of Texas system
\/}
\address{The University of Texas at Austin\\ Department of Mathematics, RLM 8.100\\ Austin, TX 78712. Email:uhlen@math.utexas.edu}

\hskip 4in \today

\begin{abstract} This is the third in a series of papers attempting to describe a uniform geometric framework in which many integrable systems can be placed.  A soliton hierarchy can be constructed from a splitting of an infinite dimensional group $L$ as positive and negative subgroups $L_\pm$ and a commuting sequence in the Lie algebra $\calL_+$ of $L_+$.  Given $f\in L_-$, there is a formal inverse scattering solution $u_f$ of the hierarchy. When there is a $2$ co-cycle on $\calL$ that vanishes on both $\calL_+$ and $\calL_-$, Wilson constructed for each $f\in L_-$ a tau function $\tau_f$ for the hierarchy. In this third paper, we prove the following results for the $n\times n$ KdV hierarchy:
\ben
\item The second partials of $\ln\tau_f$ are differential polynomials of the formal inverse scattering solution $u_f$. Moreover, $u_f$  can be recovered from the second partials of $\ln\tau_f$.
\item The natural Virasoro action on $\ln\tau_f$ constructed in the second paper is given by partial differential operators in $\ln\tau_f$.  
\item There is a bijection between phase spaces of $n\times n$ KdV hierarchy and Gelfand-Dickey (GD$_n$) hierarchy on the space of order $n$ linear differential operators on the line so that the flows in these two hierarchies correspond under the bijection.
\item Our Virasoro action on the $n\times n$ KdV hierarchy is constructed from a simple Virasoro action on the negative group. We show that it corresponds to the known Virasoro action on the GD$_n$ hierarchy under the bijection.
\een
\end{abstract}

\maketitle

\section{Introduction}

This is the third in a series of papers attempting to describe a uniform geometric framework in which many integrable systems can be placed. We defined the $n\times n$ KdV hierarchy in the first paper \cite{TerUhl11}.  We gave integral formulas for the partials of tau functions $\tau$ and a general construction of Virasoro action on $\ln\tau$ in terms of reduced frames for soliton hierarchies in the second paper \cite{TerUhl13a}. The main goal of this paper is to apply these general framework and formulas to the $n\times n$ KdV hierarchy.  In particular, we obtain the following results:
\ben
\item We can recover the formal inverse scattering solutions from partial derivatives of the tau function of the scattering data.
\item The natural Virasoro action on $\ln\tau$ explained in our previous paper \cite{TerUhl13a} is given by partial differential operators on $\ln\tau$.
\item There is a bijection between the phase space of the $n\times n$ KdV hierarchy and the space of order $n$ differential operators on the line so that flows in the $n\times n$ KdV correspond to flows in the Gelfand-Dickey (GD$_n$) hierarchy under this bijection.
\item We prove that our Virasoro action on the $n\times n$ KdV hierarchy corresponds to the well-known Virasoro action on the GD$_n$ hierarchy under the bijection. 
\een

We apologize for the fact that these results involve long and complicated computations. In the future, we would like to understand scale invariant solutions, which do not have scattering data of the sort used to construct the formulas in the paper. 

This paper is organized as follows: We set up notation and review the construction of the $n\times n$ KdV hierarchy in section 2, and review the gauge equivalence and the construction of Drinfeld-Sokolov's quotient flows in section 3. We describe cross sections of the gauge action and define the cross section flows in section 4. We prove that we can recover the formal inverse scattering solution from the second partials of $\ln\tau$ for the $n\times n$ KdV hierarchy in section 5, and  prove that the Virasoro vector fields on $\ln\tau$ are given by partial differential operators in section 6.  We show that the $n\times n$ KdV hierarchy is equivalent to the GD$_n$ hierarchy and give an algorithm to construct a solution of the $n\times n$ KdV hierarchy from a solution of the GD$_n$ hierarchy in section 7.  We prove that the Virasoro vector fields for the $n\times n$ KdV hierarchy correspond to the well-known Virasoro vector fields for the GD$_n$ hierarchy under the equivalence in the last section.

 \bs
 \section{The $n\times n$ KdV hierarchy}\label{cc}

In this section, we review the definition of the $n\times n$ KdV$_B$ hierarchy and set up some notations. Readers who have seen the first paper \cite{TerUhl11} can go directly to section \ref{gb}. 

Let $L=L(SL(n,\C))$ denote the group of smooth maps from $S^1$ to $SL(n,\C)$. 
A pair of subgroups $L_\pm$ is a {\it splitting\/} of $L$ if $L_+\cap L_-=\{e\}$ and the corresponding Lie algebra $\calL=\calL(sl(n,\C))$ will be $\calL=\calL_+\oplus \calL_-$. Let $\xi_\pm$ denote the $\calL_\pm$ component of $\xi\in \calL$ with respect to $\calL=\calL_++ \calL_-$. 

We use several splittings in this paper. The first splitting is called the {\it standard splitting\/}: 
Let $L_+(SL(n,\C))$ be the subgroup of $g\in L(SL(n,\C))$ that is the boundary value of a holomorphic map from the disk $|\l |\leq 1$, and $L_-(SL(n,\C))$ the subgroup of $f\in L(SL(n,\C))$ that is the boundary value of a holomorphic map $\ti f$ defined on $1\leq |\l |\leq \infty$ and $\ti f(\infty)=\I$.   The Lie algebras written in terms of Fourier series are given by
\begin{align*}
\calL:= \calL(sl(n,\C))&= \{\xi(\l)= \sum_j \xi_j \l^j\n \xi_j\in sl(n,\C)\},\\
\calL_+(sl(n,\C))&=\{\xi\in \calL(sl(n,\C))\,\big|\,  \xi(\l)=\sum_{j\geq 0} \xi_j \l^j\},\\
\calL_-(sl(n,\C))&=\{\xi\in \calL(sl(n,\C))\,\big|\,   \xi(\l)= \sum_{j<0} \xi_j \l^j\}.
\end{align*} 
 
Let  $N_+, N_-$ denote the subgroups of upper and lower triangular matrices in $SL(n,\C)$ with $1$ on the diagonal entries respectively, $B_+, B_-$ the subgroups of all upper and lower triangular matrices in $SL(n,\C)$ respectively, and $\calN_+, \calN_-, \calB_+, \calB_-$ the corresponding Lie subalgebras. 

 We have described the standard splitting. There is a {\it special splitting\/} 
\beq\label{ii}
\bca \calL_+= \{ \sum_{j>0} \xi_j \l^j + \xi_0\in \calL(sl(n,\C)) \n \xi_0\in \calB_+\},\\
\calL_-=\{\sum_{j<0} + \xi_0\in \calL(sl(n,\C))\n \xi_0\in \calN_-\}.\eca
\eeq
The third type of splitting uses the same $\calL_+$ as the standard splitting.
However, $\calL_-$ depends on a linear map $B:sl(n,\C)\to \calN_-$, which we described in detail when we construct the flows. 

To construct the flows we use the sequence 
\begin{align}
&\frakJ=\{J^j\n j\geq 1, j\not\equiv 0 (\mod n)\}, \label{aw2}\\
&J= e_{n1}\l + b\in \calL_+, \quad b= e_{12} + e_{23} + \ldots + e_{n-1, n}. \label{aw}
\end{align}
Next we give some properties of $J$.  A simple computation implies that
\begin{align}
&J^n=\l\I_n, \label{bg}\\
& J^i= (b^t)^{n-i} \l + b^i, \quad J^{-i}= (b^t)^i  + b^{n-i} \l^{-1}, \quad 1\leq i\leq n-1, \label{bg1}
\end{align}
where $\I_n$ is the $n\times n$ identity matrix. 

We recall a Lemma in \cite{DriSok84}. Since the proof is simple we include it. 

\blem ( \cite{DriSok84}) \label{eh} Given $\eta\in \calL(gl(n,\C))$, there exist unique $y_i\in  \calT_n$ such that $$\eta= \sum_i y_i J^i,$$  where $\calT_n$ is the sub-algebra of all diagonal matrices in $gl(n,\C)$. 
\elem 

\begin{proof}
Since $J^{nk}= \l^k \I_n$ and the formula is additive, it is enough to prove the formula for $g\in sl(n,\C)$. If $g$ is diagonal, we may choose $y_0= g= y_0\I_n$.  If $g$ is in the $j$-th diagonal, $j>0$, then we can write $g= \sum_{s=1}^{n-j}g_se_{s, j+s}$. Since $J^j= \l \sum_{i=1}^j e_{n-j+i, i} + \sum_{s=1}^{n-j} e_{s, s+j}$, we see $g= yJ^j$, where  $y=\sum_{s=1}^{n-j} g_s e_{ss}$. The proof is similar for $j<0$.
\end{proof}

  Although $J$ and diagonal matrix do not commute, there is a nice relation: 
 \beq\label{bg2} 
 J^kh = h^{\sigma^k} J^k,
 \eeq
 where $h^{\sigma^k}= \diag(h_{k+1}, \ldots, h_n, h_1, \ldots, h_k)$ if $h=\diag(h_1, \ldots, h_n)$.

The following Theorem is proved in several places in the literature (\cite{DriSok84}, \cite{TerUhl11}). 

\bthm\label{ks}
Given $u\in C^\infty(\R, \calB_-)$, there exists a unique $Q(u)\in \calL$ of the form
$$Q(u)(\l)=J+ \sum_{j\leq 0} Q_j(u) \l^j= J+\sum_{j\leq 0} y_j(u) J^j$$ satisfying 
\beq\label{ci}\bca [\p_x-(J+u), Q(u)]=0,\\
Q(u)^n=\l\I_n,\eca 
\eeq
where $y_j(u)$'s are diagonal matrices function.  
Moreover, 
\ben
\item[(i)] $Q_j(u)$ and $y_j(u)$ are differential polynomials of $u$ in $x$ variable for all $j\leq 0$,
\item[(ii)] $Q_0(u)-u\in \calN_-$.
\een 
 \ethm
 
 \begin{proof} Note that $e_{n1}z^n+ b$ is conjugate to $az$, where 
 $$\l=z^n, \quad a=\diag(1, \a, \ldots, \a^{n-1}), \quad \a=\exp(2\pi i/n).$$ So there is a unique $Q(u)$ satisfying \eqref{ci} (for a proof cf. \cite{TerUhl99b}) and entries of $Q_j(u)$'s are differential polynomials.  
 
 It remains to prove that $Q_0(u)-u\in \calN_-$. Since  $u\in \calB_-$, by \eqref{bg1}, we have $u=\sum_{i=0}^{n-1} h_{-i} J^{-i}$, where  $h_i=\sum_{j=1}^{n-i} u_{i+j, j} e_{i+j, i+j}$, where $u=(u_{ij})$. Assume that $Q(u)= J+\sum_{j\leq 0} y_j J^j$ with $y_j\in C^\infty(\R, \calT_n)$, where $\calT_n$ is the subalgebra of all diagonal matrices in $gl(n,\C)$. We use \eqref{bg2} to write \eqref{ci} as a power series in $J$ with $C^\infty(\R, \calT_n)$ coefficients on the left:
   $$\bca
   [\p_x-(J+\sum_{i=0}^{n-1} h_{-i} J^{-i}, \, J+\sum_{j< 0} y_jJ^j]=0,\\
   (J+\sum_{j\leq 0} y_jJ^j)^n=J^n.\eca $$
 Then $y_j$'s can be solved uniquely by comparing coefficients of $J^k$ for $k\leq 0$ of the above equation.  For $k=0$, we obtain $y_0= h_0$.  Theorem follows from the uniqueness of solution of \eqref{ci}. 
 \end{proof}
 
Next we consider soliton hierarchy constructed from the splitting $L_\pm$ of $L=L(SL(n,\C))$ and the {\it vacuum sequence\/} $\frakJ$ as in \eqref{aw2}. Then the flow equation generated by $J^j$ is the following evolution equation on $C^\infty(\R, Y)$, 
\beq\label{mb}
u_{t_j}= [\p_x-(J+u), (Q(u)^j)_+],
\eeq
where 
$$Y=[J, \calL_-]_+.$$
Here and henceforth we use $\xi_\pm$ to denote the $\calL_\pm$ components of $\xi\in \calL$.
$\xi=\xi_++ \xi_-$
By general theory, these flows commute. 

It can be checked that 
\beq\label{fz}
Q(u)_+= J+u
\eeq
if and only if the first flow equation is the translation $u_{t_1}= u_x$. In such case, we can identify 
$$t_1=x.$$  
But not all splittings have this last property as we show in Example \ref{dzb} that the standard splitting fails. 

\ms
\bsub {\bf The formal inverse scattering } \par

 Set $t=(t_1, \ldots, t_N)$ and
 \beq\label{kw}
  V(t)=\exp\left(\sum_{i=1}^N t_j J^j\right).
 \eeq
Given $f\in L_-$, it follows from the Local Factorization Theorem (cf. Theorem 1.2 of \cite{TerUhl11}) that there is an  subset $\calO_0$ of the origin in $\R^N$ such that we can factor 
$$V(t)f^{-1}= M(t)^{-1} E(t)\in L_-\times L_+$$
for all $t\in \calO_0$. 
It can be checked by a direct computation (cf. \cite{TerUhl11}) that 
$$u_f:= (MJM^{-1})_+ - J$$
is a solution of the hierarchy, which is called the {\it formal inverse scattering solution\/} associated to $f\in L_-$. We call $E$ and $M$ the {\it frame\/} and the {\it reduced frame\/} of $u_f$. Note that $u= 0$ is a solution of the hierarchy and $V(t)$ is a frame for $u=0$. We call $V(t)$ the {\it vacuum frame\/}.  
\esub

As mentioned before that $Q(u)_+$ is not necessary to be $J+u$ for the standard splitting. But all formal inverse scattering data solutions have this property, so we can identify $t_1$ with $x$ for these solutions:

\bprop\label{fwa}  Let $M$ denote the reduced frame of the formal inverse scattering solution $u_f$ given by $f\in L_-$. Then
\ben
\item $Q(u_f)= MJM^{-1}$ solves \eqref{ci} for $u= u_f$,
\item $Q(u_f)_+= J+ u_f$,
\item $M(\p_x-J)M^{-1}= \p_x-(J+ u_f)$,
\item $(u_f)_{t_1}= (u_f)_x$, and we may identify $t_1$ with $x$ for formal inverse scattering solutions.
\een
\eprop

\begin{proof}
It follows from the construction of $u_f$ that we have $M= EfV^{-1}$.  Since $V$ and $J$ commute, we get 
$ MJM^{-1}= EfJf^{-1}E^{-1}$ and
$$M_{t_1}M^{-1}= E_{t_1} E^{-1}-EfJ f^{-1} E^{-1}= E_{t_1}E^{-1} -MJM^{-1}.$$
But $E_{t_1}E^{-1}\in \calL_+$ and $M_{t_1}M^{-1}\in \calL_-$ imply that $$(MJM^{-1})_+= E_{t_1}E^{-1},$$
\beq\label{hw}
M_{t_1}M^{-1}= -(MJM^{-1})_-.
\eeq  
Statements of the Proposition follow from \eqref{hw} and a straight forward computation. 
\end{proof}

There is a sequence of natural commuting flows on $L_-$ constructed from the splitting $L_\pm$ and the vacuum sequence $\frakJ$. Solutions of these flows on $L_-$ give rise to solutions of the soliton flows \eqref{mb} (cf. \cite{TerUhl99b}):

\bthm \label{ht} Let $J=e_{n1}\l+b$ be as in \eqref{aw}.  If $M(t)\in L_-$ satisfies 
$$M_{t_j} M^{-1}= -(MJ^jM^{-1})_-, \quad 1\leq j\leq N,$$ then $u:=(MJM^{-1})_+ - J$ is a solution of the hierarchy constructed from $L_\pm$ and the vacuum sequence $\frakJ=\{J^j\n j\not\equiv 0 \, (\mod n)\}$. 
\ethm

 We give below three known KdV type hierarchies arising from splittings discussed in the beginning of the section. 

\beg \label{mw} {\it The $n\times n$ KdV$_B$ hierarchy} (\cite{TerUhl11})\par

Let $B:sl(n,\C)\to \calN_-$ be a linear map satisfying 
\ben
\item[(a)] $\calB_+\subset \Ker(B)$,
\item[(b)] $B$ satisfies 
\beq\label{ee}
 [B(\xi), B(\eta)]= B([B(\xi), \eta]+ [\xi, B(\eta)])
 \eeq
 for all $\xi, \eta\in sl(n,\C)$,
\item[(c)] $\dim([J, \calL_-^B]_+)= n-1$. 
\een
Note that $B$ satisfies \eqref{ee} if and only if
\beq\label{ed}
\calL^B_-=\{\xi\in \calL\n \xi(\l)= B(\xi_{-1}) +\sum_{i<0} \xi_i \l^i\}.
\eeq
is a Lie subalgebra of $\calL(sl(n,\C))$.   
Let $\calL_+=\calL_+(sl(n,\C))$ as in the standard splitting and $\calL_-=\calL_-^B$. Then $\calL_\pm$ is a splitting of $\calL(sl(n,\C))$. The hierarchy constructed from the splitting $\calL_\pm$ and the vacuum sequence  $\frakJ$ (as in \eqref{aw2}) is called the {\it $n\times n$ KdV$_B$ hierarchy\/} (cf. \cite{TerUhl11}).  Note that 
$$Y:=[J, \calL_-^B]_+\subset \calB_-.$$
  \eeg
 
 \ss
\beg\label{ke} {\it The $n\times n$ KdV hierarchy}\,\, \cite{TerUhl11}\par

Let 
\beq\label{ka}
\L= \sum_{k=1}^{n-1} \frac{1-\a^k}{1-\a} e_{k+1, k}, \quad \a= e^{\frac{2\pi i}{n}}.
\eeq
A simple computation implies that $\{\L^i b^{n-1}\L^j\n 0\leq i, j\leq n-1\}$ is a basis of $sl(n,\C)$ and $\{\L^i b^{n-1} \L^j\n i, j\geq 0, i+j < n-1\}$ is basis for the subalgebra $\calN_+$.
It was proved in \cite{TerUhl11} that the linear map
$B:sl(n)\to \calN_-$ defined by  
\beq\label{gg} 
B(\L^i b^{n-1}\L^j)= \L^i b^t \L^j
\eeq
satisfies conditions (a)-(c) given in Example \ref{mw}. 
  We call the hierarchy constructed from the splitting $(\calL_+(sl(n,\C)), \calL_-^B)$ given by this $B$ and the vacuum sequence $\frakJ$ as in \eqref{aw2} {\it the $n\times n$ KdV hierarchy\/}. The flows in this hierarchy are evolution equations on $C^\infty(\R, Y)$, where 
\beq\label{mz}
Y=[J, \calL_-^B]_+=\left\{\sum_{i=1}^{n-1} u_i \L^i\,\bigg|\, u_i\in \C\right\}.
\eeq

  When $n=2$, the only $B$ which satisfies \eqref{ee} is $B(e_{12})= \pm e_{21}$. The $2\times 2$ KdV hierarchy is the KdV hierarchy and  the flow generated by $J^3$ is the KdV flow
  $$q_t=\frac{1}{4}(q_{xxx} \pm 6 qq_x).$$
  
  When $n=3$, the flow generated by $J^2$ in the $3\times 3$ KdV hierarchy is the following coupled non-linear Schr\"odinger equations:
$$
\bca 
(u_1)_t= -\frac{\sqrt{3}}{3}\, i\,  (u_1)_{xx} + \frac{3-\sqrt{3}\, i}{3} (u_2)_x,\\
(u_2)_t = \frac{\sqrt{3}}{3}\, i\, (u_2)_{xx} + 2 u_1(u_1)_x.
\eca
$$
\eeg

\beg\label{en} {\it The $n\times n$ mKdV hierarchy} (Drinfeld-Sokolov \cite{DriSok84})
 
 Let $\calL_\pm$ be the special splitting of $\calL(sl(n,\C))$ defined by \eqref{ii}. 
The $n\times n$ modified KdV hierarchy is the hierarchy constructed from the splitting $\calL_\pm$ of $\calL(sl(n,\C))$ and the vacuum sequence $\frakJ$ (as in \eqref{aw2}). 
 The flows in this hierarchy are evolution equations for maps $u=\diag(u_1, \ldots, u_n)$ with $\sum_{i=1}^n u_i=0$.  
\eeg

\beg\label{dzb} 
Let $\calL_\pm=\calL_\pm(sl(n,\C))$ be the standard splitting of $\calL=$ $\calL(sl(n,\C))$. The flows in the hierarchy constructed from $\calL_\pm$ and the vacuum sequence  $\frakJ$ as ub \eqref{aw2} are evolution equations on $C^\infty(\R, Y)$, where 
$$Y=[J, \calL_-]_+= \C(e_{11}-e_{nn}) \oplus( \oplus_{i=2}^{n-1} (\C e_{i1} \oplus \C e_{n,i})).$$
 
 By Proposition \ref{fwa}, $Q(u_f)_+ = J+ u_f$ for formal inverse scattering solutions.  But 
for general $u\in C^\infty(\R, Y)$, the solution $Q(u)$ of \eqref{ci} need not satisfy the condition
$Q(u)_+= J+u$. A direct but long computation shows that
$$u=u_0(e_{11}-e_{nn}) +  u_{n-1} e_{n1} +  \sum_{i=2}^{n-2} (u_{i-1} e_{i1} + v_{i-1} e_{n, n-i+1}) \in C^\infty(\R, Y)$$
 satisfies $Q(u)_+= J+u$ if and only if $u_i, v_i$'s  satisfy a system of $n-1$ ordinary differential equations of the form
$$\bca v_i= c_i u_i + p_i, & 1\leq i\leq n-2,\\
 u_{n-1}= p_{n-1},\eca$$
where $p_i$ is some differential polynomial of $u_0, u_1, \ldots, u_{i-1}$ in $x$ variable for $1\leq i\leq n-1$. The flows \eqref{mb} leave such a $u$ invariant and the restriction gives a system of partial differential equations for $u_0, \ldots, u_{n-2}$.  
For example, for $n=2$, $u=\bpm u_0&0\\ u_1& -u_0\epm$ satisfies $Q(u)= J+u$ if and only if
$$u_1= (u_0)_x- u_0^2.$$
The flow generated by $J^3$ for such $u$ gives the following evolution equation for $u_0$:
$$(u_0)_t= \frac{1}{4}((u_0)_{xxx}- 6 ((u_0)_x)^2).$$
Note that if $u_0$ is a solution of the above equation then $q= 2(u_0)_x$ is a solution of the KdV, i.e., $q_t= \frac{1}{4}(q_{xxx}- 6 qq_x)$.  
\eeg

\bs

\section{The gauge group action and quotient flows}\label{gb}

We review the construction of Drinfeld-Sokolov's KdV type hierarchy associated to the Kac-Moody algebra $A_{n-1}^{(1)}$ (the {\it DS $A_{n-1}^{(1)}$-KdV hierarchy}). They show these flows are equivalent to the Gelfand-Dickey flows on the space of $n$-th order differential operators on the line.  We give a different proof later. 

Let $B_\pm, N_\pm, \calB_\pm, \calN_\pm$, and $\calG_k$ be as in section \ref{cc}, and $J$ as in \eqref{aw}.  
Given an integer $k$, define a bi-linear form $\li\, , \ri_k$ on $\calL=\calL(sl(n,\C))$ as follows:
\beq\label{aa}
\li \xi, \eta\ri_{k} = \sum_j \tr(\xi_j \eta_{-j+k}),
\eeq
 the coefficient of $\l^{k}$ of $\tr(\xi(\l)\eta(\l))$, where $\xi(\l)=\sum_j \xi_j\l^j$ and $\eta(\l)= \sum_j \eta_j\l^j$. 
 
 Let $S$ be a linear subspace of $\calB_-$. Define 
 \beq\label{gk}
 \calM(S):=\{L_u=\p_x-(J+u)\n u\in C^\infty(\R, \calS)\}.
 \eeq
The group $C^\infty(\R, N_-)$ acts on $C^\infty(\R, \calL)$ by point-wise conjugation,
\beq\label{hx}
(g, \xi)\mapsto  g \xi g^{-1}, \quad g\in N_-, \xi\in \calL,
\eeq
and further acts on the space of connections 
$$\{ \p_x-\xi\n \xi\in C^\infty(\R, \calL)\}$$ by gauge transformation,
$$(g, \p_x -\xi)\mapsto g(\p_x-\xi) g^{-1} = \p_x -(g \xi g^{-1}+ g_x g^{-1}).$$
 
Note that for $g\in C^\infty(\R, N_-)$ and $\xi, \eta\in C^\infty(\R, \calL)$, we have
\begin{subequations}
\begin{gather}
(g \xi g^{-1})_+= g \xi_+ g^{-1}, \label{hy1}\\
\li g \xi g^{-1}, g \eta g^{-1}\ri_k = \li \xi, \eta\ri_k, \label{hy2}
\end{gather}
\end{subequations}  
where $\xi_\pm$ is the projection of $\xi$ onto $\calL_\pm$. 

\blem The gauge action of $C^\infty(\R, N_-)$ on the space of connections,
$\{\p_x-\xi\n \xi\in C^\infty(\R, \calL)\}$, leaves the space
$$\calM(\calB_-)=\{L_u= \p_x -(J+u)\n u\in C^\infty(\R, \calB_-)\}$$
invariant.
\elem

\begin{proof}
Given $u\in C^\infty(\R, \calB_-)$ and $\D\in C^\infty(\R, N_-)$, we note that
\begin{align*}
\D L_u \D^{-1}&= \p_x -(\D (J+u)\D^{-1} + \D_x \D^{-1})\\ 
&= \p_x-(\D e_{n1} \D^{-1}\l + \D b \D^{-1} + \D u\D^{-1} + \D_x\D^{-1}\\
&= \p_x -(e_{n1}\l + b + (\D b\D^{-1}-b) + \D u\D^{-1}+ \D_x\D^{-1})\\
&= \p_x-(J + (\D b\D^{-1}-b) + \D u\D^{-1}+ \D_x\D^{-1}).
\end{align*}
Some facts to note in the calculation are 
$$[e_{n1},\calN_-]=0, \quad \D b\D^{-1}-b\in \calN_-, \quad \D u\D^{-1}\in\calB_-.$$ 
\end{proof}

Given $u\in C^\infty(\R, \calB_-)$ and $\D\in C^\infty(\R, N_-)$, define 
$$\D\ast u:= \D_x \D^{-1}+ \D u\D^{-1} + \D b\D^{-1}-b.$$  Then we have
\beq\label{gc}
\D L_u \D^{-1}= L_{\D \ast u}.
\eeq

\bcor\label{kv} The tangent space to the gauge group $C^\infty(\R, N_-)$ orbit at $L_u$ in $\calM(\calB_-)$ is 
$$\{[L_u, \xi]=\xi_x -[b+u, \xi]\n \xi\in C^\infty(\R, \calN_-)\}.$$
\ecor

\bcor\label{ku} Let $u\in C^\infty(\R, \calB_-)$, and $Q(u)= J+\sum_{i\leq 0} Q_i(u)\l^i$ be as in Theorem \ref{ks}. Then
\ben
\item $Q_0(u)-u\in \calN_-$,
\item $Q(\D\ast u) = \D Q(u)\D^{-1}$. 
\een
\ecor

\begin{proof} (1) was proved in Theorem \ref{ks} (i). 

It follows from $[L_u, Q(u)]=0$ and $\D L_u \D^{-1}= L_{\D \ast u}$ that we have 
$$\D [L_u, Q(u)]\D^{-1}= [L_{\D\ast u}, \D Q(u) \D^{-1}]=0.$$ By the uniqueness part of Theorem \ref{ks}, we get $Q(\D \ast u) = \D Q(u) \D^{-1}$. 
\end{proof}

\bdefn  For each $j\not\equiv 0$ (mod $n$), define the $j$-th flow on $C^\infty(\R, \calB_-)$ by
\beq\label{hz}
\p_{t_j} L_u= \frac{\p u}{\p t_j} =[L_u, (Q(u)^j)_+],
\eeq
where $\xi_+=\sum_{i\geq 0} \xi_i \l^i$ if $\xi=\sum_i \xi_i \l^i$.  \edefn

To see that \eqref{hz} defines a flow on $C^\infty(\R, \calB_-)$, we note that $[L_u, Q(u)]=0$ implies $[L_u, Q(u)^j]=0$. So  we have 
$$[L_u, (Q(u))^j_+]=-[L_u, (Q(u)^j)_-].$$ The left-hand side of the above equality lies in $\calL_+(sl(n,\C))$ and the right hand side is of the form 
$$[e_{n1}, sl(n,\C)] + \sum_{j<0} \xi_j \l^j.$$
Since $[e_{n1}, sl(n,\C)]\subset \calB_-$, we have $[L_u, (Q(u)^j)_+]\in C^\infty(\R, \calB_-)$. Hence \eqref{hz} is a flow equation on $C^\infty(\R, \calB_-)$. 

Next we show that flow \eqref{hz} commute with the gauge action.

\blem\label{gm} Suppose  $u_0, \ti u_0\in C^\infty(\R, \calB_-)$ such $L_{\ti u_0} = \D L_{u_0} \D^{-1}$ for some $\D\in C^\infty(\R, N_-)$.  If $u(x,t_j)$ is the solution of \eqref{hz} with $u(x,0)= u_0(x)$, then 
$$\ti u:=\D\ast u=  \D_x\D^{-1} + (\D b\D^{-1}-b) + \D u\D^{-1}$$
is the solution of \eqref{hz} with $\ti u(x,0)= \ti u_0(x)$, or equivalently
$$\p_{t_j} L_{\ti u}= [L_{\ti u}, (Q(\ti u)^j)_+].$$
\elem 

\begin{proof} Note that $L_{\ti u} = \D L_u \D^{-1}$.  By Corollary \ref{ku},  $Q(\ti u)= Q(\D \ast u)= \D Q(u) \D^{-1}$. But 
$(\D \xi^j\D^{-1})_+=\D (\xi^j)_+\D^{-1}$ for $\xi\in C^\infty(\R, \calL)$. Hence we have
\begin{align*}
\p_{t_j} L_{\ti u} &= \p_{t_j} \D(x) L_u \D(x)^{-1}= \D(x) (\p_{t_j} L_u) \D(x)^{-1}\\
&= \D(x) [L_u, Q(u)^j_+] \D(x)^{-1}= [\D L_u\D^{-1}, \D (Q(u)^j)_+\D^{-1}]\\
&= [L_{\ti u}, (\D Q(u)\D^{-1})^j_+]= [L_{\ti u}, (Q(\ti u))_+]. 
\end{align*}
\end{proof}

The proof of Lemma \ref{gm} also gives

\bcor \label{gm1}  Let $\D\in C^\infty(\R, N_-)$, and $u, v\in C^\infty(\R, \calB_-)$ such that $L_v= \D L_u \D^{-1}$, and $Q(u), Q(v)$ be solutions of \eqref{ci} for $u$ and $v$ respectively. Then $\D Q(u) \D^{-1}= Q(v)$. 
\ecor

In general $\p_{t_1}L_u\not=\p_x L_u$, but they lie in the same gauge orbit:

\bprop\label{jt}
The flow $\p_{t_1}$ is equivalent to the flow given by $\p_x$ on $\calM(\calB_-)$ under the gauge group $C^\infty(\R, N_-)$.
\eprop

\begin{proof}
We use the same notation as in Theorem \ref{ks}. By Corollary \ref{ku} (i), $Q_0(u)= u+ v$ with some $v\in C^\infty(\R, \calN_-)$. 
Compute directly to get
\begin{align*}
\p_{t_1} L_u&= [L_u, Q(u)_+]= [\p_x-(b+u), b+Q_0(u)]=  [\p_x-(b+u), b+ u+v] \\
&= u_x + [L_u, v].
\end{align*}
 By Corollary \ref{kv}, $(\p_{t_1}-\p_x)u$ is tangent to the gauge orbit. 
\end{proof}

From Lemma \ref{gm} and Proposition \ref{jt} we get the main theorem of this section:

\bthm (\cite{DriSok84}) The flow \eqref{hz} on $\calM(\calB_-)$ induces a well-defined quotient flow on the orbit space (or the quotient) $\calM(\calB_-)/C^\infty(\R, N_-)$.  Moreover, we have $\p_{t_1}= \p_x$ on the orbit space.  
\ethm

\bs

\section{Cross section flows}\label{gd}

We use the same notation as in section \ref{gb}: For $u\in C^\infty(\R, \calB_-)$, 
$$L_u= \p_x- (J+u),$$
and for a linear subspace  $S$ of $\calB_-$, 
$$\calM(S)=\{L_v=\p_x-(J+v)\n v\in C^\infty(\R, S)\}.$$
Let $S_0$ and $S_1$ be linear subspaces of $\calB_-$ defined by
\begin{align}
S_0&= \oplus_{i=1}^{n-1} \C \L^i, \label{gh0}\\
S_1&= \oplus_{i=1}^{n-1} \C e_{n, n-i}, \label{gh1}
\end{align}
where $\L$ is given by \eqref{ka}. Note that $\calM(S_0)$ is the phase spaces of the $n\times n$ KdV hierarchy.

Recall that $C^\infty(\R, N_-)$ acts on $\calM(\calB_-)$ by  
$$g(\p_x-(J+u))g^{-1}= \p_x-(J + gug^{-1} + g_x g^{-1}).$$
In this section, we give a sufficient condition for a linear subspace $S$ of $\calB_-$ such that 
$\calM(S)$ is a cross section for the $C^\infty(\R, N_-)$ gauge action on $\calM(\calB_-)$, i.e.,  every $C^\infty(\R, N_-)$-orbit in $\calM(\calB_-)$ meets $\calM(S)$ exactly once.  We also give an algorithm to compute the induced cross section flows from \eqref{hz}. 

\bprop Suppose $S\subset \calB_-$ is a linear subspace such that $\calM(S)=\{L_u=\p_x-(J+u)\n u\in S\}$ is a cross-section of the gauge group $C^\infty(\R, N_-)$ on $\calM(\calB_-)$. Then the flows \eqref{hz} induce flows on the cross section. Moreover, if $S$ and $\ti S$ are two different cross-sections, the induced flows on $\calM(S)$ are gauge equivalent to the induced flows on $\calM(\ti S)$.
\eprop

\begin{proof} Since $\calM(S)$ is a cross-section of the gauge group, there exists a projection $\pi:\calM(\calB_-)\to \calM(S)$ such that $\pi \circ g= \pi$ for all $g\in C^\infty(\R, N_-)$. 
To obtain the flow on $\calM(S)$, we simply project the flow on $\calM(\calB_-)$ onto $\calM(S)$ using the gauge group.  It is more or less a tautology that the induced flows on $\calM(S))$ and $\calM(\ti S)$ are gauge equivalent. They will look very different as evolution equations, but will be gauge equivalent. 
\end{proof}

Dirnfeld and Sokolov shows that $\calM(S_1)$ is a {\it cross section\/} of the  action of $C^\infty(\R, N_-)$ on $\calM(\calB_-)$, where $S_1$ is defined by \eqref{gh1}.  Below we include a proof of this because the proof also gives rise to a condition on affine subspace $S$ such that $\calM(S)$ is a cross section. 

  First we set up some notation. Let 
  \beq\label{gp}
  \calG_i=\oplus_{\ell-k=i} \C e_{k\ell}.
  \eeq  Then 
$[\calG_i, \calG_j]\subset \calG_{i+j}$, $\calG_i=0$ if $|i|\geq n$.  
For $y=(y_{ij})\in sl(n,\C)$, let 
\beq\label{gpa}
\tr_k(y)= \sum_{j-i=k} y_{ij}.
\eeq

\bprop  (\cite{DriSok84}) \label{ba}
Given $u\in C^\infty(\R,\calB_-)$, there exist unique $\D\in C^\infty(\R, N_-)$ and $\xi=\sum_{j=1}^{n-1} \xi_j e_{n, n-j}\in C^\infty(\R, S_1)$  such that $\D L_u \D^{-1}=L_\xi$, i.e., 
\beq\label{ax}
\D (\p_x-(J+u)) \D^{-1}= \p_x- (J+ \xi).
\eeq
Moreover,  let $u=\sum_{i=0}^{n-1} u_{-j}$ with $u_{-j}\in \calG_{-j}$, and $\xi=\sum_{j=1}^{n-1} \xi_j e_{n, n-j}$. Then
\beq\label{bw}
\xi_j= \tr_{-j}(u_{-j})+\eta_j, \quad 1\leq j\leq n-1,
\eeq
where $\eta_j$ is a polynomial in $u_0, u_{-1}, \ldots, u_{-(j-1)}$ and their $x$ derivatives.
\eprop

\begin{proof}
Since $\D e_{n1}= e_{n1}\D= e_{n1}$, \eqref{ax} is equivalent to
\beq\label{az}
[\D, b] = -\D u - \D_x + \xi \D.
\eeq
  It is easy to check that $\ad(b):\calG_{-(j+1)}\to \calG_{-j}$ is injective and the image is 
\beq\label{ay}
[b, \calG_{-(j+1)}]= \{\eta=\sum_{i=1}^{n-i} \eta_i e_{j+i, i}\in \calG_{-j}\n \tr_{-j}(\eta):= \sum_{i=1}^{n-j} \eta_i=0\}.
\eeq
Write $\D= \sum_{j=0}^{n-1} \D_{-j}$ and $u=\sum_{j=0}^{n-1} u_{-j}$ with $\D_{-j}, u_{-j}\in \calG_{-j}$.  
Because $\D\in N_-$, we have $\D_0=\I$.  Equate the $\calG_0$ component of \eqref{ay} to get $[\D_{-1}, b]=-u_0$. Since $\tr(u_0)=0$, we have $\D_{-1}= \ad(b)^{-1}(u_0)$.  

We will prove the remainder of this Proposition by induction.  Suppose we have solved $\xi_1, \ldots, \xi_{j-1}$ and $\D_{-1}, \ldots, \D_{-j}$.  
Note that $\xi_0=0$.  Equate the $\calG_{-j}$ component in \eqref{az} to get
\beq\label{kb}
[\D_{-(j+1)}, b]= -\sum_{i=0}^j \D_{-(j-i)} u_{-i} - (\D_{-j})_x + \sum_{i=0}^j \xi_i e_{n, n-i}\D_{-(j-i)}.
\eeq
The above equation is solvable for $\D_{-(j+1)}$ if and only if the $\tr_{-j}$ of the right-hand side is zero,
i.e.,
\beq\label{kc}
\tr_{-j}\left(-\sum_{i=0}^j \D_{-(j-i)} u_{-i} -(\D_{-j})_x + \sum_{i=0}^{j-1} \xi_i \D_{-(j-1)}\right)+ \xi_j=0.
\eeq  
This gives a formula for $\xi_j$ in terms of $\xi_1, \ldots, \xi_{j-1}$ and $\D_{-1}, \ldots, \D_{-j}$. Since $\ad(b)$ is injective on $\calG_{-(j+1)}$, we can solve for $\D_{j+1}$. 
\end{proof} 

\bdefn\label{oa}
Given $L_u=\p_x-(J+u)\in \calM(\calB_-)$, let $S_1$ be as in \eqref{gh1}, $\xi\in C^\infty(\R, S_1)$ and $\D\in C^\infty(\R, N_-)$ be as in Proposition \ref{ba} such that $\D L_u \D^{-1}= L_\xi$  or equivalently, \eqref{ax} holds.  Define $\Psi:\calM(\calB_-)\to \ti \calM(S_1)$ by
$$\Psi(\p_x-(J+u))= \p_x-(J+\xi).$$
We call $\xi$ the {\it GD variable\/} of $u$. 
\edefn

\ss
The proof of Proposition \ref{ba} gives a simple criterion on subspace $S$ such that $\calM(S)$ is a  cross sections of the $C^\infty(\R, N_-)$ gauge action on $\calM(\calB_-)$:

\bcor \label{dzs} Let $S=\oplus_{i=1}^{n-1} \C v_{-i}$ be a linear subspace of $\calB_-$ with  $v_{-i}\in \calG_{-i}$ for $1\leq i\leq n-1$.  If  
$$\tr_{-j}(v_{-j})\not=0, \quad {\rm for \, all\,\, } 1\leq j\leq n-1,$$ then $\calM(S)$  is a cross section of the $C^\infty(\R, N_-)$ action on $\calM(\calB_-)$. Hence  the map $\Psi$ defined by \eqref{oa} maps $\calM(S)$ onto $\calM(S_1)$ isomorphically, where  $S_1=\oplus_{i=1}^{n-1} \C e_{n, n-i}$ as in \eqref{gh1}.
 \ecor  
 
 We proved in a previous paper \cite{TerUhl11} that $\tr_{-j}(\L^j)\not=0$ (cf. Lemma 2.7 of \cite{TerUhl11}).  So $\calM(\calS_0)$ is a cross section, where $S_0$ is given by \eqref{gh0}.

\bcor\label{dz} Let $S_0$ and $S_1$ be as defined by \eqref{gh0} and \eqref{gh1}. Then
\ben
\item The restriction of $\Psi:\calM(\calB_-)\to  \calM(S_1)$  to $\calM(S_0)$ is a bijection,
\item  if $\Psi(L_u)= L_\xi$ with $u=\sum_{j=1}^{n-1} u_j \L^j$ and $\xi= \sum_{j=1}^{n-1} \xi_j e_{n, n-j}$, then
$$\xi_j= \tr_{-j}(\L^j)u_j +\eta_j,$$
where $\eta_j$ is a polynomial in $u_1, \ldots, u_{j-1}$ and their $x$ derivatives.
\een 
\ecor

We also need the following corollary later. 

\bcor\label{ob} Given $\eta\in C^\infty(\R, B_-)$ and $L_\xi \in \calM(S_1)$, if $\eta L_\xi \eta^{-1}\in \calM(S_1)$, then $\eta=c\I_n$ for some constant $c$. 
\ecor

\begin{proof}
Suppose $\eta L_\xi \eta^{-1}= L_{\ti \xi}\in \calM(S_1)$.  Then $\eta (J+\xi) \eta^{-1}+ \eta_x \eta^{-1}= \ti \xi$. Hence we have
\beq\label{aza}
[b,\eta] = \eta \xi + \eta_x - \ti \xi \eta.
\eeq
Write $\eta=\sum_{i=0}^{n-1}\eta_{-i}$ with $\eta_{-i}\in \calG_{-i}$.  Equate the $\calG_1$ component of \eqref{aza} implies that $[\eta_0, b]=0$. Hence $\eta_0= h \I_n$ for some function $h$. Equate the $\calG_0$ component of \eqref{aza} to get 
$[b,\eta_{-1}]= h_x\I_n$.  The left hand side has trace zero, so $h_x=0$. Hence $h$ is a constant $c$.  
\end{proof}

\bsub{\bf The cross-section flow}\par

  If $\calM(S)$ is a cross section of the $C^\infty(\R, N_-)$ gauge action on $\calM(\calB_-)$, then we can compute the induced cross section flow on $\calM(S)$ as follows: By Proposition \ref{ba}, given $v\in C^\infty(\R, S)$, there is a unique $\eta_j(v)\in C^\infty(\R, \calN_-)$ such that 
 \beq\label{gna}
 [L_v, (Q(v)^j)_+-\eta_j(v)]\in C^\infty(\R, S),
 \eeq
  where $\xi_+$ is the projection with respect to the standard splitting of $\calL$ and $Q(v)$ is the solution of \eqref{ci} for $u=v$. 
  In fact, $\eta_j(v)$ can be solved algebraically from \eqref{gna} and entries $\eta_j$ are polynomials in $v\in C^\infty(\R, S)$ and its $x$ derivatives.   The induced {\it cross section flow on $\calM(S)$\/} is
\beq\label{gn} 
\p_{t_j}v = [L_v, (Q(v)^j)_+-\eta_j(v)].
\eeq 
   
    We call the hierarchy of cross section flows \eqref{gn} on $\calM(S_1)$ the {\it DS $A_{n-1}^{(1)}$-KdV hierarchy\/}.
Although cross section flows on various cross sections may look different, they are equivalent under the gauge group. 

In general, the cross section flows \eqref{gn} do not come from a splitting. In particular, the DS $A_{n-1}^{(1)}$-KdV hierarchy on $\calM(\calS_1)$ does not come from a splitting. But the cross section flows on $\calM(S_0)$ do come from a splitting, where $S_0$ is defined by \eqref{gh0}.  In fact, for $u\in C^\infty(\R, S_0)$, we have $(Q(u)^j)_+-\eta_j(u)=\pi_+(Q(u)^j)$, where $\pi_+$ is the projection of $\calL$ with respect to the splitting of $n\times n$ KdV hierarchy given in Example \ref{ke}. 
\esub

\bs

\section{Tau functions\/}\label{mq}
 
Let $\li\, , \ri_k$ denote the bilinear form on $\calL$ defined by \eqref{aa}. 
The usual choice of cocycle is (cf. \cite{PreSeg86}) 
\beq\label{ab} 
w(\xi, \eta)=\li \xi_\l, \eta\ri_{-1}= \sum_j j\tr(\xi_j \eta_{-j}).
\eeq
We will use $\xi_\l$ to denote $\p\xi/\p\l$.  

In this section, we assume that $\calL_\pm$ is  a splitting of $\calL=\calL(sl(n,\C))$ such that  
\beq\label{ab2}
\calL_+\subset \{ \sum_{i\geq 0} \xi_i \l^i\}, \quad \calL_-\subset \{\sum_{i\leq 0} \xi_i \l^i \n \xi_0\in \calN_-\}.
\eeq
Then $\li \calL_+, \calL_+\ri_{-1}=0$ and $\calL_\pm$ is compatible with the $2$-cocyle $w$ defined by \eqref{ab}, i.e., $w(\xi, \eta)=0$ for all $\xi, \eta\in \calL_+$ and for all $\xi, \eta\in \calL_-$. 
We consider hierarchy constructed from such splitting and the sequence $\frakJ=\{J^j\n j\not\equiv 0 (\mod n)\}$, where 
 $J=e_{n1}\l +b$ as in \eqref{aw}. Note that $[J, \calL_-]_+\subset \calB_-$, where $\calB_-$ is the space of lower triangular matrices in $sl(n,\C)$.  It turns out that all the splittings we have used in section \ref{cc} satisfy conditions (a) and (b). 
  
 It was proved in \cite{TerUhl13a} that
 tau functions, which are complex valued functions of $t=(t_1, \ldots, t_N)$, are defined for the hierarchy constructed from $\calL_\pm$ and the vacuum sequence $\frakJ$.  The following integral formulas for partials of $\ln\tau_f$ are proved in  \cite{TerUhl13a}: 
\begin{align}
&(\ln\tau_f)_{t_j}= \li MJ^jM^{-1}, M_\l M^{-1}\ri_{-1}, \label{csa}\\
&(\ln\tau_f)_{t_1t_j}= \li MJ^jM^{-1}, (J)_\l\ri_{-1} =\li MJ^j M^{-1}, e_{n1}\ri_{-1}, \label{csb}\\
& (\ln\tau_f)_{t_jt_k}= \li MJ^j M^{-1}, \p_\l (MJ^k M^{-1})_+\ri_{-1}.\label{ag3}
\end{align}
Here $M$ is the reduced frame for the formal inverse scattering solution  $u_f$ given by $f\in L_-$.  Note that formulas similar to these for other systems appear in many places as definition of tau functions, including the work of \cite{AraLeu03}. 

  We will prove the following two results in this section:
 \ben
 \item[\bu] $(\ln\tau_f)_{t_it_j}$ is invariant under the $C^\infty(\R, N_-)$ gauge transformations and only depends on the GD variable $\xi$ of $u_f$ (cf. Definition \ref{oa}). In particular, $(\ln\tau_f)_{t_it_j}$ is independent of the splittings and $\rd(\ln \tau)$ is defined for DS $A_{n-1}^{(1)}$-KdV hierarchy, where $\rd$ is the differential with respect to $t_1, \ldots, t_N$. 
  \item[\bu] We can recover $u_f$ from $\{(\ln\tau_f)_{t_1t_j}\n 1\leq j\leq n-1\}$ for the $n\times n$ KdV hierarchy. 
 \een
  
 The map $f\mapsto u_f$ is not injective. In fact, if $h\in L_-$ satisfying $hJ=Jh$, then $u_f= u_{fh}$ and the reduced frame $\ti M$ for $u_{fh}$ is equal to $Mh$, where $M$ is the reduced frame for $u_f$. By \eqref{csa}, we see that  $(\ln\tau_{fh})_{t_j}=(\ln\tau_f)_{t_j}+ c_j$, where $c_j=\li J^j, h_\l h^{-1}\ri_{-1}$ is a constant. But the second partials of $\ln\tau_f$ and $\ln\tau_{fh}$ are the same. 
  
Next we prove that $(\ln\tau_f)_{t_j t_k}$ only depends on the GD variables of $u_f$. So they are gauge invariant. 

\bprop\label{ge} Let $L_\pm$ be a splitting of $\calL=\calL(sl(n,\C))$ satisfying \eqref{ab2},  $J=e_{n1}\l+b$ as in \eqref{aw},  $f\in L_-$, and $M$ and $\xi$ the reduced frame and GD variable of the formal inverse scattering solution $u_f$ of the hierarchy constructed from $\calL_\pm$ and $\frakJ=\{J^j\n j\not\equiv 0\, (\mod n)\}$ respectively. Then
\ben
\item[(i)] $(\ln\tau_f)_{t_1t_j}= \li Q(\xi)^j, e_{n1}\ri_{-1}$,
\item[(ii)] $(\ln\tau_f)_{t_jt_k} =\li Q(\xi)^j, \p_\l (Q(\xi)^k)_+)\ri_{-1}$,
\een
where $Q(\xi)$ is the solution of \eqref{ci} for $\xi$. 
\eprop

\begin{proof} Let $L_{u_f}= \p_x-(J+u_f)$. By definition of the GD variable (Definition \ref{oa}), there exists a unique $\D(t)\in N_-$ independent of $\l$ such that $L_\xi = \D L_u \D^{-1}$. By Proposition \ref{fwa}, $L_{u_f}= M(\p_x-J)M^{-1}$ and $Q(u_f)= MJM^{-1}$ is the solution of \eqref{ci} for $u_f$. So $L_\xi = \D L_u \D^{-1} = \D M(\p_x-J) M^{-1}\D^{-1}$, which implies that 
$[\p_x-(J+\xi), \D MJM^{-1}\D^{-1}] = 0$. Hence 
$$Q(\xi)= \D M J M^{-1}\D^{-1}$$
is the solution for \eqref{ci} for $\xi$. 
Since $\D$ is independent of $\l$, we have 
$$\D \p_\l((Q^k(u_f))_+)\D^{-1}= \p_\l(\D Q^k(u_f)\D^{-1})_+.$$
 Use \eqref{ag3} and the above equations to see the following.
 \begin{align*}
 &(\ln\tau_f)_{t_jt_k}= \li Q(u_f)^j, \p_\l (Q(u_f)^k)_+\ri_{-1} \\
 &\quad = \li \D Q(u_f)^j \D^{-1}, \D \p_\l (Q(u_f)^k)_+ \D^{-1}\ri_{-1} = \li Q(\xi)^k, \p_\l(Q(\xi)^j)_+)\ri_{-1}.
 \end{align*}
 This proves that $(\ln\tau_f)_{t_jt_k}$ only depends on the GD variable $\xi$ of $u_f$. 
 \end{proof}
 
We need explicit formula for the first $n$ terms of the solution $Q(\xi)$ of \eqref{ci} as a power series in $J$ for $\xi\in C^\infty(\R, S_1)$. 

\blem\label{od}  Let $Q(\xi)= J + \sum_{i=0}^\infty h_{-i} J^{-i}$ be the solution of \eqref{ci} for $\xi=\sum_{i=1}^{n-1}\xi_j e_{n, n-j}$.  Then $h_0=0$ and
\beq\label{hf}
 h_{-j}= \frac{1}{n} \xi_j \bc + \zeta_j, \quad 1\leq j\leq n-1,
 \eeq
  $\bc= \diag(1, \ldots, 1, -(n-1))$, and $\zeta_j$ is a polynomial of $\xi_1, \ldots, \xi_{j-1}$ and their $t_1$ derivatives.  
\elem

\begin{proof} We use \eqref{bg1} and a simple computation to show that
 \beq\label{go1} 
\xi= \sum_{i=1}^{n-1} \xi_i e_{nn} J^{-i}.
\eeq  
Write $Q(\xi)= J+\sum_{i\leq 0} Q_i(u)\l^i$. 
 By Corollary \ref{ku}, $Q_0(\xi)- \xi \in \calN_-$. So $Q_0(\xi)\in \calN_-$. It follows from \eqref{bg1} that $Q(\xi)$ as a power series of $J$ has the form 
  \beq\label{hc}
 Q(\xi)= J+ \sum_{i=1}^\infty h_{-i} J^{-i},
 \eeq
Note that the coefficient of $J^0$ in the expansion of $Q(\xi)$ is zero. 
 
 Since $Q(\xi)$ is the solution of \eqref{ci}, we get
 \beq\label{bm}\bca
 [\p_x-(J+ \sum_{i=1}^{n-1} A_{-i} J^{-i}), \, J+ \sum_{i<0} h_i J^i]=0, \\
 (J + \sum_{i<0} h_i J^i)^n= \l \I_n,
 \eca
 \eeq
 where  $A_i= \xi_i e_{nn}$ for $1\leq i\leq n-1$.  
  Write 
 $$h_{-j}= \diag(h_{-j1}, \ldots, h_{-jn}).$$
 We will prove \eqref{hf} by induction:
 For $j=1$, compare coefficient of $J^0$ in the first equation of \eqref{bm} and use \eqref{bg2} to get
 $$[J, h_{-1}J^{-1}] + [A_{-1}J^{-1}, J]=0.$$
 By \eqref{bg2}, we get
 $$h_{-1}^\sigma - h_{-1}= A_{-1}^\sigma - A_{-1}.$$
 This implies that 
 \beq\label{bn}
 h_{-11}= \ldots = h_{-1, n-1}, \quad h_{-11}- h_{-1n}= \xi_1.
\eeq
Compare coefficient of $J^{n-2}$ in the second equation of \eqref{bm} and use \eqref{bg2} to get
$$h_{-1} + h_{-1}^\sigma + \cdots + h_{-1}^{\sigma^{n-1}}=0.$$
So we have
\beq\label{boa}
\sum_{i=1}^n h_{-1i}=0.
\eeq  
Solve \eqref{bn} and \eqref{boa} to get 
$h_{-1}= \frac{\xi_1}{n}\bc$, where $\bc= \diag(1, \ldots, 1, -(n-1))$.

Next we define weights as follows:  
$$w(\xi_i)= i, \quad w((\xi_i)^{(j)})= i+j, \quad w(\zeta_1\zeta_2)= w(\zeta_1)+ w(\zeta_2),$$
where $\xi^{(j)}=\p_x^j \xi$. 
Fix $1\leq i < n-1$, assume that for $1\leq j \leq i$ we have 
$$h_{-j}= \frac{\xi_j}{n} \bc + \zeta_j$$
and the weight $w(h_j)=j$,
where $\zeta_j$ depends on $\xi_1, \ldots, \xi_{j-1}$. 
 Compare coefficients of $J^{-i}$ and $J^{n-i-2}$ of the first and the second equations of \eqref{bm} respectively and the induction hypothesis to get
\begin{align*}
&h_{-(i+1)}^\sigma- h_{-(i+1)} = A_{-(i+1)}^\sigma - A_{-(i+1)} + \phi_i,\\
&\sum_{k=0}^{n-1}h_{-(i+1)}^{\sigma^k}=\psi_i,
\end{align*}
where $\phi_i$ and $\psi_i$ are functions of $\xi_1, \ldots, \xi_i$ and their $t_1$ derivatives and the weights of $\phi_i$ and $\psi_i$ are $i+1$.  These two equations imply that 
$$h_{-(i+1)} = \frac{\xi_{i+1}}{n} \bc + \zeta_{i+1},$$
where $\bc= \diag(1, 1, \ldots, -(n-1))$ and $\zeta_{i+1}$ is a function of $\xi_1, \ldots, \xi_i$ and their $t_1$ derivatives.  This proves \eqref{hf} for $j=i+1$.  
\end{proof}
 
    The following Theorem shows that we can recover the GD variable of $u_f$ from $\{(\ln\tau_\xi)_{t_1t_j}\n 1\leq j\leq n-1\}$.  
      
\bthm \label{hh} We use the same notation as in Proposition \ref{ge}. Let $\xi=\sum_{i=1}^{n-1} \xi_i e_{n, n-i}$ denote the GD variable of $u_f$.
  Then
 \beq\label{ha}
(\ln \tau_f)_{t_1t_j}= \frac{j}{n} \xi_j + y_j, \quad 1\leq j\leq n-1,
\eeq
where $y_j$'s are polynomials in $\xi_1, \ldots, \xi_{j-1}$ and their $t_1$ derivatives. 
\ethm

\begin{proof} By Proposition \ref{fwa}, the solution $Q(u_f)$ for \eqref{ci} is $MJM^{-1}$, where $M$ is the reduced frame for $u_f$.  
It follows from Proposition \ref{ge} that we have
$$(\ln\tau_f)_{t_1t_j}= \li Q(\xi)^j, e_{n1}\ri_{-1}.$$

If $y$ is a diagonal matrices, then it follows from \eqref{bg1} that  we have
$$\li yJ^i, e_{n1}\ri_{-1} =\bca  0, & {\rm if\,\, } i\not=-1,\\
 \tr(y e_{11}), & {\rm if\,\,} i=-1.\eca
 $$
  By Lemma \ref{od}, $Q(\xi)= J+\sum_{i=1}^\infty h_{-i}J^{-i}$ with $h_{-i}$ given by \eqref{hf} for $1\leq i\leq n-1$.  
Hence
$$Q(\xi)^j=\sum_{i\leq j} k_i J^i =  (J+ \sum_{i<0} h_i J^i)^j.$$ Therefore we can compute $k_i$ in terms of $h_k$'s.  
 In particular, we have
 $$k_{-1}= \sum_{i=0}^{j-1} J^{j-1-i} h_{-j} J^{-j} J^i+ \b_j(h_{-1}, \ldots, h_{-(j-1)}).$$
   This proves that 
   \begin{align*}
    (\ln\tau_f)_{t_1t_j} &=\li k_{-1}, e_{n1}\ri \\
  & = \li \sum_{i=0}^{j-1}  h_{-j}^{\sigma^{j-i-1}} J^{-1}, e_{n1}\ri + p_j (h_{-1}, \ldots, h_{-(j-1)}).
 \end{align*}
The first term is equal to $\sum_{i=1}^j h_{-ji}$,  where $h_j=\diag(h_{j1}, \ldots, h_{jn})$.   Hence we have
 \beq\label{he}
  (\ln\tau_f)_{t_1t_j}= \sum_{i=1}^j h_{-ji} + p_j(h_{-1}, \ldots, h_{-(j-1)}).
 \eeq 
For $1\leq j\leq n-1$, we use Lemma \ref{od} to see that
$$\sum_{i=1}^j h_{-ji} =\frac{j}{n} \xi_j + \g_j,$$
where $\g_j$ is a function of $\xi_1, \ldots, \xi_{j-1}$ and their $t_1$-derivative. So formula \eqref{ha} follows. 
\end{proof}

As a consequence of Corollary \ref{dzs}, Proposition \ref{ba} and Theorem \ref{hh}, we have the following.

\bcor  Let $\calL_\pm$ and $J$ as in Proposition \ref{ge}. Suppose $[J, \calL_-]_+=\oplus_{i=1}^{n-1} \C v_i$ with $v_i\in \calG_i$ and  $\tr_{-i}(v_i)\not=0$ for all $1\leq i\leq n-1$, where $\calG_k$ and $\tr_k$ are defined by \eqref{gp} and \eqref{gpa} respectively.   Then given $f\in L_-$, we can recover the formal inverse scattering solution $u_f$ from $\{(\ln\tau_f)_{t_1t_i}\n 1\leq i\leq n-1\}$.  
\ecor

\bcor We can recover the formal inverse scattering solution $u_f$ with scattering data $f$ from $\{(\ln\tau_f)_{t_1 t_j}\n 1\leq j\leq n-1\}$ for the $n\times n$ KdV hierarchy. 
\ecor

\bs
\section{Virasoro action on tau functions}\label{mr}

In this section we assume $\calL=\calL(sl(n,\C))$, and $\calL_\pm$ is a splitting of $\calL$ such that $\calL_+= \calL_+(sl(n,\C))$ as in the standard splitting and $\calL_-\subset \calN_- + \calL_-(sl(n,\C))$.  Then $\calL_\pm$ satisfies \eqref{ab2} given in the beginning of \ref{mq}. The main goal is to prove that the Virasoro vector fields constructed in \cite{TerUhl13a} on $\ln\tau_f$ of the hierarchy given by $\calL_\pm$ and $\frakJ=\{J^j\n j\not\equiv 0 \, (\mod n)\}$ are partial differential operators of $\ln\tau_f$.  Here $J=e_{n1}\l +b$ as defined by \eqref{aw}.

Let $V(t)=\exp(\sum_{j=1}^N t_j J^j)$ be the vacuum frame defined by \eqref{kw}. Here we assume $$t_{kn}=0, \quad k\geq 1.$$   Given $f\in L_-$, factor $V(t)f^{-1}=M(t)^{-1}E(t)$ with $M(t)\in L_-$ and $E(t)\in L_+$.  Recall that the formal inverse scattering solution 
$$u_f=(MJM^{-1})_+-J= E_{t_1}E^{-1}- J$$ and $M$ is the reduced frame of $u_f$.  

The positive Virasoro algebra $\cv_+$ is the Lie algebra $\calV_+=\{\xi_\ell\n \ell \geq -1\}$ with the bracket relations 
$$[\xi_j, \xi_k]= (k-j) \xi_{j+k}, \quad \forall\,\, j, k\geq -1.$$
An action of $\calV_+$ on $N$ is a sequence of tangent vector fields $X_j$ on $N$ satisfies
$$[X_j, X_k]=(k-j) X_{j+k}, \quad j, k\geq -1.$$
We call these $X_j$ {\it Virasoro vector fields on $N$\/}. 

Set 
 \beq\label{dm4}
  \Xi=\frac{1}{n}\diag(0, 1, \ldots, (n-1)).
 \eeq
We proved in \cite{TerUhl13a} that
\beq\label{ap}
\d_\ell(f) = -(\l^\ell(\l f_\l f^{-1}+ f\Xi f^{-1}))_- f 
\eeq
are Virasoro vector fields on $L_-$ and the induced Virasoro vector fields on reduced frames and $\ln\tau_f$ are
\begin{align}
\d_\ell (M) M^{-1} &= -(\l^\ell E(\l f_\l f^{-1}+ f\Xi f^{-1})E^{-1})_-, \label{ms1}\\
\d_\ell(\ln\tau_f) &= \li \l^\ell E(\l f_\l f^{-1} + f\Xi f^{-1}) E^{-1}, \l E_\l E^{-1}\ri_0.\label{ms2}
 \end{align}

  Henceforth in this section we use the following notations:
\begin{align}
&\G f= \l f_\l + f\Xi, \label{dm2} \\
&\hat \G f= \l f_\l + [f, \Xi],  \label{dm3}
\end{align}
where $\Xi$ is defined by \eqref{dm4}. Note that $\hat \G$ is a derivation, i.e.,
\beq\label{fs}
\hat \G(f_1f_2)= (\hat \G f_1) f_2 + f_1\hat \G(f_2).
\eeq
The {\it degree\/} of $\xi\in \calL(sl(n,\C))$ is $k$ if $\xi(\l)= \sum_{i\leq k} \xi_i \l^i$.

\ss

We need the following two lemmas to write the formula \eqref{ms1} in a better form. Both lemmas can be proved by direct computations.

\blem \label{cv} Let $\hat \G$ be defined by \eqref{dm3}, $J=e_{n1}\l+b$ as in \eqref{aw}, and $V(t)$ the vacuum frame defined by \eqref{kw}.  Then we have
\begin{align}
&\hat \G J= \frac{1}{n} J, \quad \hat \G J^k= \frac{k}{n} J^k, \label{kx}\\
&\hat \calJ:= (\hat \G V)V^{-1}=\frac{1}{n}\sum_{k=1}^N kt_k J^k,  \label{cx}\\
& (\l^\ell \hat \G V(t))V(t)^{-1}=\l^\ell \hat\calJ=\frac{1}{n} \sum_{k=1}^N k t_k J^{n\ell+k}, \label{db}
\end{align}
where $t_{nk}=0$.
\elem

\blem   Let $E$ and $M$ be the frame and reduced frame for the formal inverse scattering solution $u_f$. Then we have
\beq\label{cf}
E((\G f) f^{-1}) E^{-1}+ \l E_\l E^{-1} = (\G M) M^{-1} + M\hat\calJ M^{-1},
\eeq
where $\hat\calJ$ is defined by \eqref{cx}.
\elem

\bthm\label{fb} 
 The Virasoro vector fields \eqref{ms1} on the reduced frame $M$ is given by
 \beq\label{fc}
 \d_\ell(M) M^{-1}= -(\l^\ell (\G M) M^{-1} + \l^\ell M\hat \calJ M^{-1})_-.
 \eeq
 Here $\hat \calJ$ is defined by \eqref{cx}, $\G$ is defined by \eqref{dm2}, and $\ell\geq -1$.
\ethm

Since $u_f=(MJM^{-1})_+-J$, we have
$$\d_\ell u_f= -[(\d_\ell M) M^{-1}, MJM^{-1}]_+.$$
Note that $\Xi$ is diagonal the diagonal entries of $J^m$ are zero if $m\not\equiv 0 (\mod n)$. So $\li M\Xi M^{-1}, MJ^{n\ell +k}M^{-1}\ri_{-1}= 0$ if $k\not\equiv 0\, (\mod n)$.  This implies the following.

\bcor The $\calV_+$-action on $u_f$ is given by
$$\d_\ell u_f= -[(\l^{\ell+1} M_\l M^{-1})_-, J+u]_+ - \sum_{ k\not\equiv 0 (\mod n)} \frac{kt_k}{n} [(MJ^{n\ell+k} M^{-1})_-, J+u]_+.$$
\ecor

Note that entries of $Q(u_f)=MJM^{-1}$ are differential polynomials of $u_f$, but entries of $M_\l M^{-1}$ are not.  So the Virasoro vector fields on $u_f$ are not given by differential operators. 

\bthm\label{dg}
The Virasoro vector fields \eqref{ms2} on $\calX= \ln\tau_f$ are given by the following formulas:
\begin{align*}
&\d_\ell\calX = \frac{1}{n} \sum_{k\geq 1} kt_k \calX_{t_{n\ell+k}} + \frac{1}{2n}\sum_{k=1}^{n\ell-1} \left(\calX_{t_k}\calX_{t_{n\ell-k}} + \calX_{t_k t_{n\ell-k}}\right) \\
&\qquad \quad  +(\frac{1}{2n} -\frac{1}{2}) c_\ell(f), \quad \ell\geq 1,\\
&\d_0\calX = \frac{1}{n} \sum_{k\geq 1} k t_k \calX_{t_k} +\frac{1}{2} c_0(f),\\
&\d_{-1}\calX =\frac{1}{n} \sum_{k>n}  k t_k \calX_{t_{k-n}} +\frac{1}{2n} \sum_{k=1}^{n-1} k(n-k) t_k t_{n-k},
\end{align*}
where $c_\ell(f)= \li \l^\ell ((\G f) f^{-1})^2\ri_0$ and $\G$ is the operator defined by \eqref{dm2}.  Here we assume $\calX_{t_{nk}}=0$ for all $k\geq 1$. 
\ethm

We need several Lemmas to prove this theorem. 

\blem\label{ft}
Let $M$ denote the reduced frame of the formal inverse scattering solution $u_f$, and $Q:= Q(u_f)= MJM^{-1}$. Then $Q^n=\l \I_n$, $[Q^i, Q^j]=0$, and
\begin{align}
& \l (Q^j)_\l= [(\G M)M^{-1}, Q^j] + \frac{j}{n} Q^j, \quad 1\leq j\leq n-1, \label{cy}\\
& \tr(\l (Q^j)_\l Q^{n-j})= j\l, \quad 1\leq j\leq n-1. \label{cz}
\end{align}
\elem

\begin{proof} 
By \eqref{bg1}, $J^j= (b^t)^{n-j} \l + b^j$. So
$$Q^j= MJ^j M^{-1} = M((b^t)^{n-j} \l + b^j)M^{-1}.$$
Compute directly to get
\begin{align*}
\l (Q^j)_\l&= [\l M_\l M^{-1}, Q^j] + \l M(b^t)^{n-j} M^{-1} \\
&= [\G M M^{-1}- M\Xi M^{-1}, Q^j] + \l  M(b^t)^{n-j} M^{-1} \\
&= [\G M M^{-1}, Q^j] + M(-[\Xi, (b^t)^{n-j}\l + b^j]+ \l (b^t)^{n-j})M^{-1}.
\end{align*}
A simple computation implies that
$$[\Xi, (b^t)^{n-i}]= \frac{n-i}{n} (b^t)^{n-i}, \quad [\Xi, b^i]= -\frac{i}{n} b^i.$$ 
So \eqref{cy} follows.  

We use \eqref{cy}, $\tr([\xi_1, \xi_2]\xi_3)= \tr(\xi_1[\xi_2, \xi_3])$, and $[Q^i, Q^j]=0$ to compute
\begin{align*}
\tr(\l (Q^j)_\l, Q^{n-j}) &= \tr(([(\G M) M^{-1}, Q^j]+ \frac{j}{n} Q^j)Q^{n-j})\\& = \tr((\G M) M^{-1}[Q^j, Q^{n-j}]) + \frac{j}{n} \tr (Q^n) = j\l.
\end{align*}
This gives \eqref{cz}.
\end{proof}

\blem\label{gf} Let $M$ denote the reduced frame of the formal inverse scattering solution $u_f$ with scattering data $f\in L_-$.  
If $k\geq 0$, then 
$$\li (\G M) M^{-1}, \l^k\ri_0=\li \l^k (\G f) f^{-1}\ri_0.$$
\elem

\begin{proof}
By \eqref{cf}, we get
\begin{align*}
\li &\l^k (\G M) M^{-1}\ri_0= \li \l^k(E(\G f) f^{-1}E^{-1} + \l E_\l E^{-1} - M\hat \calJ M^{-1})\ri_0\\
&= \li\l^k (\G f) f^{-1}\ri_0 + \li \l ^{k+1}E_\l E^{-1}\ri_0 - \li \l^k \hat\calJ\ri_0,
\end{align*}
where $\hat \calJ$ is given by \eqref{cx}. 
The degree of $\l E_\l E^{-1}$ is $1$, so the second term is zero. If $k\geq 1$, then third term is zero. If $k=0$, then the third term is equal to
$$\li \hat\calJ\ri_0=\frac{1}{n} \sum_{k=1}^{n-1} k t_k \li J^k\ri_0 =0. $$
\end{proof}

\blem Let $M$ be the reduced frame for $u_f$, and $Q=MJ^jM^{-1}$.  If $1\leq i\leq n-1$ and $\eta\in \calL$, then
\beq\label{dd}
\li \p_\l \eta, \p_\l(Q^i)_+\ri_{-1}=0.
\eeq
\elem

\begin{proof}
Since $J^i= (b^t)^{n-i}\l + b^i$ for $1\leq i\leq n-1$, $Q^i$ has degree $1$ in $\l$. So $(Q^i)_+=\xi_0$ does not depend on $\l$.  But $\p \eta$ has no $\l^{-1}$ term. 
\end{proof}

\blem
Let $M$ be the reduced frame for $u_f$, $P=(\G M) M^{-1}$. Then for $0\leq j\leq n-1$, we have
\beq\label{dc}
\tr(P^2 \l)= \tr(PQ^j PQ^{n-j})-\frac{1}{2} \tr([P, Q^j][P, Q^{n-j}]).
\eeq
\elem

\begin{proof}
A direct computation implies that
\begin{align*}
\tr([P, Q^j][P, Q^{n-j}])&= 2\tr(PQ^jPQ^{n-j})- 2\tr(Q^n P^2)\\
&= 2\tr(PQ^jPQ^{n-j})- 2\tr(\l P^2).
\end{align*}
\end{proof}

Note that for $j=0$, the right hand side of \eqref{dc} gives $\tr(P^2Q^n)=\tr(P^2\l)$. 

\blem\label{dk}
Let $S, T\in \calL$, and $\p=\p_\l$. Then
\begin{align}
& \p S_+ = (\p S)_+, \quad \p S_-= (\p S)_-, \label{bk3}\\
&\li \p (ST)\ri_{-1}=\li \p S, T\ri_{-1} +\li S, \p T\ri_{-1}=0, \label{bk4}\\
&\li \p S, T\ri_{-1}=\li \p S_+, T\ri_{-1} + \li \p S, T_+\ri_{-1},\label{bk2}\\
&\li \p(\l S)_+, \p T\ri_{-1}= \li \p S_+, \p(\l T)\ri_{-1}+ 2\li S_+, \p T\ri_{-1}.\label{bk1}
\end{align}
\elem

\bsub {\bf Proof of Theorem \ref{dg}} \hfil

By \eqref{ms2}, we have
$$\d_\ell \calX =\li \l^\ell E(\G f)f^{-1} E^{-1}, \l E_\l E^{-1}\ri_0.$$
We use \eqref{cf},  
$$\tr (\xi\eta)=\frac{1}{2} \tr( (\xi+\eta)^2)-\tr(\xi^2)-\tr(\eta^2)),$$
and $\li \l^{\ell+2}(E_\l E^{-1})^2\ri_0=0$ for $\ell\geq -1$ to get
 \begin{align*}
& \d_\ell\calX= \frac{1}{2} \li \l^\ell ((\G M) M^{-1}+ M\hat \calJ M^{-1})^2\ri_0-\frac{1}{2} \li \l^\ell ((\G f) f^{-1})^2\ri_0\\
 &= \frac{1}{2} \li \l^\ell ((\G M) M^{-1})^2\ri_0 + \frac{1}{2} \li \l^\ell \hat \calJ^2 \ri_0 + \li \l^\ell (\G M) M^{-1}, M\hat \calJ M^{-1}\ri_0 - \frac{1}{2} c_\ell(f)\\
 &=\frac{1}{2} (I)+\frac{1}{2} (II)+(III) -\frac{1}{2} c_\ell(f),
  \end{align*}
 where $c_\ell(f)=\li \l^\ell ((\G f) f^{-1})^2\ri_0$ and $\hat \calJ$ as in \eqref{cx}.  Note that 
 $$c_{-1}(f)=0, \quad c_0(f)= \li \Xi^2\ri_0 =\frac{1}{n^2}\sum_{i=1}^{n-1} i^2.$$ 
 
 First we compute $(II)$.   Recall that  $J^k=(b^t)^{n-k}\l + b^k$ for $1\leq k\leq n-1$,
and $J^n=\l \I_n$. 
Since the degree of $\hat J^2$ is zero, we have $(II)=0$ if $\ell>0$.  For $\ell=0$, since the constant term of the expansion of $\hat J^2$ in $\l$ is in $\calN_+$, $(II)=0$.  Hence we have
 $$(II)=0, \quad \ell\geq 0.$$ 
 For $\ell=-1$, we have $(II)= \li \l^{-1} \hat \calJ^2\ri_0= \li \l^{-1} \sum_{k=1}^{n-1} \frac{k(n-k)}{n^2} t_kt_{n-k} J^n\ri_0$.
 This implies that 
 $$(II) =\frac{1}{n} \sum_{k=1}^{n-1} k(n-k) t_k t_{n-k}, \qquad {\rm if\,\, } \ell=-1.$$
   
   To compute (III), we write
   $$(III)= \li \l^\ell (\l M_\l M^{-1}), \sum_{k\geq 1} k t_k MJ^kM^{-1}\ri_0 +\li \l^\ell M\Xi M^{-1}, M\hat\calJ M^{-1|}\ri_0.$$
   Note that if $k\not\equiv 0 (\mod n)$, then 
   \beq\label{fe}\li M\Xi M^{-1}, MJ^k M^{-1}\ri_0=\li \Xi, J^k\ri_0=0.
   \eeq
   So we have
   \beq\label{fd}
   \li \l^\ell M\Xi M^{-1}, M\hat \calJ M^{-1}\ri_0=\li \l^\ell \Xi, \hat \calJ\ri_0=0.
   \eeq
   Hence
   \begin{align*}
   (III) &= \li \l^\ell (\l M_\l M^{-1}), \sum_k k t_k M J^k M^{-1}\ri_0 \\
   &= \li \l M_\l M^{-1}, \sum_k kt_k MJ^{k+n\ell} M^{-1}\ri_0.
   \end{align*}
   It follows from $(\ln\tau_f)_{t_j}= \li M J^j M^{-1}, \l M_\l M^{-1}\ri_0$ that we have
  $$(III)= \frac{1}{n} \sum_{k\geq 1} k t_k \calX_{t_{n\ell + k}}, \quad {\rm if \,\,} \ell\geq 0.$$
 For $\ell=-1$,  we have
 \begin{align*} (III)&= \li M_\l M^{-1} +\l^{-1}M\Xi M^{-1}, M\hat\calJ M^{-1}\ri_0\\
 &= \li M_\l M^{-1}, M\hat\calJ M^{-1}\ri_0+ \li \l^{-1}\Xi, \hat \calJ\ri_0
= \li M_\l M^{-1}, M\hat\calJ M^{-1}\ri_0 \\
&= \li M_\l M^{-1}, \frac{1}{n} \sum_{k=1}^{n-1} kt_k MJ^kM^{-1}\ri_0\\
& \quad+\li M_\l M^{-1}, \frac{1}{n} \sum_{k>n} kt_k MJ^kM^{-1}\ri_0\\
&= (i)+ (ii).
\end{align*}
 It follows from $J^n=\l \I_n$ that
 $$(ii)= \frac{1}{n} \sum_{k>n} kt_k \calX_{t_{k-n}}.$$
 The degree of $M_\l M^{-1}$ is $-2$ and the degree of $MJ^k M^{-1}$ is $1$ for $1\leq k\leq n-1$. So we have $(i)=0$. This proves that 
 $$(III)= \frac{1}{n} \sum_{k>n}k t_k \calX_{t_{k-n}}, \qquad {\rm if\,\,} \ell=-1.$$ 
 
 Next we compute (I). 
 For $\ell=-1$, since the degree of  $\l^{-1} ((\G M) M^{-1})^2$ in $\l$ is $-1$, we have 
 $$(I)=0, \quad {\rm\,\, } \ell=-1.$$
 
 If $\ell=0$, then 
 \begin{align*}
 (I)&= \li ((\G M) M^{-1})^2\ri_0= \li (\l M_\l M^{-1} + M\Xi M^{-1})^2\ri_0\\
 & = \li (\l M_\l M^{-1})^2\ri_0 + \li \Xi^2\ri_0 + 2\li \l M_\l M^{-1}, M\Xi M^{-1}\ri_0.
 \end{align*}
  Since the degrees of the first term and the third term in $\l$ are $-2$ and $-1$ respectively, we have
 $$(I)= \li \Xi^2\ri_0 = \frac{1}{n^2} \sum_{k=1}^{n-1}k^2 = c_0(f), \quad {\rm if\,\,} \ell=0.$$
 
To compute $(I)$ for $\ell\geq 1$, we set $Q=MJM^{-1}$. Then $Q^j= MJ^j M^{-1}$ and $Q^n=\l\I_n$.
 Write 
 $$P= (\G M) M^{-1}=\l M_\l M^{-1} + M\Xi M^{-1}.$$ Then
 \begin{align*}
 (I)&= \li \l^\ell P^2\ri_0 = \li \l^{\ell-1}\l P^2\ri_0 =\li \l^{\ell-1} Q^n P^2\ri_0\\
 &=\frac{1}{n}\sum_{i=0}^{n-1} \li \l^{\ell-1} Q^i Q^{n-i} P^2\ri_0.
  \end{align*}
A direct computation implies that
\beq
\tr([P, \xi][P, \eta])= 2\tr(P\xi P\eta) -\tr(P^2(\xi\eta+\eta\xi)).
\eeq
Let $\xi= Q^i$ and $\eta= Q^{n-i}$ in the above equation to see that 
 \begin{align*}
 (I)&= \frac{1}{n}\sum_{i=0}^{n-1} \li \l^{\ell-1} P Q^i PQ^{n-i}\ri_0 - \frac{1}{2n} \li \l^{\ell-1} [P, Q^i][P, Q^{n-i}]\ri_0\\
 &= \frac{1}{n} (A) -\frac{1}{2n} (B).
 \end{align*}
 Let $\li\eta\ri_k=$ the coefficient of $\l^k$ of $\tr(\eta(\l))$.  
 Note that the degrees of $P$ and $PQ^i$ and $PQ^{n\ell-i}$ are $0$, $1$ and $\ell$ respectively, where $1\leq i\leq n-1$ So we have
\begin{align*}
(A)&= \sum_{i=0}^{n-1} \li PQ^i PQ^{n\ell-i}\ri_0\\
& =\sum_{i=0}^{n-1}\left(\li PQ^i\ri_1\li PQ^{n\ell-i}\ri_{-1} +\sum_{k=0}^{\ell} \li PQ^i\ri_{-k} \li PQ^{n\ell-i}\ri_k\right).
\end{align*}
Since $P Q^i= (\l M_\l+ M\Xi)((b^t)^{n-i} \l + b^i)M^{-1}$, 
$$\li PQ^i\ri_1= \li \Xi, (b^t)^{n-i}\ri_0= 0, \quad 1\leq i\leq n-1.$$
Hence we have 
\begin{align*}
(A)&= \sum_{i=0}^{n-1}\sum_{k=0}^{\ell} \li PQ^i\ri_{-k} \li PQ^{n\ell-i}\ri_k =\sum_{i=0, k=0}^{n-1, \ell} \li P Q^i \l^k\ri_0 \li PQ^{n\ell-i}\l^{-k}\ri_0.
\end{align*}
Use $Q^n=\l\I$ to continue the computation and we get
$$(A) = \sum_{i=0, k=0}^{n-1, \ell} \li P Q^{nk+i}\ri_0 \li PQ^{n\ell-(nk+i)}\ri_0= \sum_{j=0}^{n\ell}\li PQ^j\ri_0 \li PQ^{n\ell-j}\ri_0.$$
By \eqref{csa}, $\calX_{t_j}= \li Q^j, \l M_\l M^{-1}\ri_0$. So 
$$\li Q^j, P\ri_0=\li Q^j, \l M_\l M^{-1} + M\Xi M^{-1}\ri_0= \li J^j, \Xi \ri_0+ \calX_{t_j}.$$
But $\li J^j, \Xi\ri_0=0$. Hence $\li Q^j, P\ri_0= \calX_{t_j}$ for $j\not\equiv 0 (\mod n)$.  Substitute this into $(A)$ above to get
\begin{align*}
(A)&= \sum_{k=0}^{\ell} \li PQ^{nk}\ri_0 \li PQ^{n(\ell-k)}\ri_0 +\sum_{j=1, \, j\not\equiv 0 (\mod n)}^{n\ell-1} \calX_{t_j} \calX_{t_{n\ell -j}} \\
&=\sum_{k=0}^{\ell} \li P\l^k\ri_0\li P\l^{\ell-k}\ri_0 +\sum_{j=1, \, j\not\equiv 0 (\mod n)}^{n\ell-1} \calX_{t_j} \calX_{t_{n\ell -j}}
\end{align*}

By Lemma \ref{gf}, we have
\begin{align*} \sum_{k=0}^\ell\li P\l^k\ri_0\li P\l^{\ell-k}\ri_0 &= \sum_{k=0}^\ell \li \G f)f^{-1}\ri_{-k} \li(\G f) f^{-1}\ri_{-(\ell-k)}\\
&=\li ((\G f) f^{-1})^2\ri_{-\ell}=\li \l^\ell (\G f) f^{-1})^2\ri_0= c_\ell(f).
\end{align*}

Thus we get
$$(A)= \sum_{j=1,\, j\not\equiv 0 (\mod n)}^{n\ell-1}\calX_{t_j} \calX_{t_{n\ell-j}} + c_\ell(f).$$

Next we compute 
\begin{align*}
(B)&= \sum_{i=0}^{n-1} \li \l^{\ell-1} [P, Q^i] [P, Q^{n-i}]\ri_0, \quad {\rm by\,\, \eqref{cy},} \\
&= \sum_{i=0}^{n-1} \li \l^{\ell-1}(\l (Q^i )_\l- \frac{i}{n} Q^i)(\l (Q^{n-i})_\l -\frac{n-i}{n} Q^{n-i})\ri_0\\
&= \sum_{i=0}^{n-1} \li \l^{\ell-1} \l (Q^i)_\l \l (Q^{n-i})_\l\ri_0 -\frac{i}{n} \li \l^{\ell-1} Q^i \l ( Q^{n-i})_\l\ri_0\\
&\qquad - \frac{n-i}{n} \li \l^{\ell-1} \l( Q^i)_\l Q^{n-i}\ri_0 + \frac{i(n-i)}{n^2} \li Q^{n\ell}\ri_0.\\
\end{align*}
By \eqref{cz}, the second and third terms are zero. The fourth term is $\frac{i(n-i)}{n} \d_{\ell,0}$. 
So for $\ell\geq 1$,  we get
\begin{align*}
(B)&= \sum_{i=0}^{n-1} \li \l^{\ell-1} \l( Q^i)_\l \l (Q^{n-i})_\l\ri_0= \sum_{i=0}^{n-1} \li \l^{\ell+1} \p_\l Q^i \p_\l Q^{n-i}\ri_0\\
&= \sum_{i=0}^{n-1} \li \l^\ell \p_\l Q^i \p_\l Q^{n-i}\ri_{-1} = \sum_{i=0}^{n-1} \li \p_\l(\l^\ell Q^i) - \ell \l^{\ell-1} Q^i, \p_\l Q^{n-i}\ri_{-1}.
\end{align*}
Use \eqref{cz} to get 
$$\li \l^{\ell-1} Q^i, \p_\l Q^{n-i}\ri_{-1}= (n-i)\li \l^{\ell-1}\I_n\ri_0,$$
which is zero if $\ell \geq 1$.  So we have
\begin{align*}
(B)&= \sum_{i=0}^{n-1} \li \p(\l^\ell Q^i), \p Q^{n-i}\ri_{-1}= \li \p (\l^\ell), \p\l\ri_{-1} +\sum_{i=1}^{n-1} 
\li \p (\l^\ell Q^i), \p Q^{n-i}\ri_{-1}\\
&= \sum_{i=1}^{n-1} \li \p (\l^\ell Q^i), \p Q^{n-i}\ri_{-1}.
\end{align*}
Apply \eqref{bk2}  to the above equation to get 
$$(B)=\sum_{i=1}^{n-1} \li \p(\l^\ell Q^i)_+, \p Q^{n-i}\ri_{-1} + \li \p(\l^\ell Q^i), \p (Q^{n-i})_+\ri_{-1}$$
But for $1\leq j\leq n-1$, $Q^j$ has degree $1$, so $\p_\l Q^i$ is independent of $\l$. But $\p_\l \xi$ has no $\l^{-1}$ term.  Hence the second term of the right hand side of $(B)$ given above is zero and we get
$$(B)=\sum_{i=1}^{n-1} \li \p(\l^\ell Q^i)_+, \p Q^{n-i}\ri_{-1}.$$
Apply  \eqref{bk1} $\ell$ times and use \eqref{dd} to get
\begin{align*}
(B)&= \sum_{i=1}^{n-1} \li \p (Q^i)_+, \p (\l^\ell Q^{n-i})\ri_{-1} + 2 \sum_{k=1}^\ell \li (\l^{\ell-k}Q^i)_+, \p (\l^{k-1} Q^{n-i})\ri_{-1}\\ 
&= 2\sum_{i=1}^{n-1} \sum_{k=1}^\ell \li (\l^{\ell-k}Q^i)_+, \p (\l^{k-1} Q^{n-i})\ri_{-1}\\
&=2\sum_{i=1, k=1}^{n-1, \ell} \li (Q^{n(\ell-k)+i})_+, \p (Q^{nk-i})\ri_{-1} \\
&= -2\sum_{i=1, k=1}^{n-1, \ell} \li \p (Q^{n(\ell-k)+i})_+, (Q^{nk-i})\ri_{-1}.
\end{align*}
By \eqref{ag3}, we have $\calX_{t_jt_k}= \li Q^j, \l \p_\l (Q^k)_+\ri_0.$ Therefore
$$(B)= -2\sum_{i=1, k=1}^{n-1, \ell}\calX_{t_{n\ell-{nk-i}} t_{nk-i}}=-2\sum_{j=1}^{n\ell-1} \calX_{t_jt_{n\ell-j}}. $$
So we have 
$$(I)= \frac{A}{n}- \frac{B}{2n} = \frac{1}{n}\left( \sum_{j=1, j\not\equiv 0}^{n\ell -1} \calX_{t_j}\calX_{t_{n\ell-j}} + \calX_{t_j t_{n\ell-j}} + c_\ell(f)\right), \quad {\rm if\,\, } \ell\geq 1.$$
Collecting terms to get the formulas for $\d_\ell \calX$.
\qed
\esub

\brem The splittings of the $n\times n$ KdV$_B$ hierarchy satisfy \eqref{ab2}.  So the Virasoro vector fields on tau functions for the $n\times n$ KdV$_B$ hierarchy are given by partial differential operators, which are independent of the choice of $B$ modulo constants.  
\erem

\bs
\section{Equivalence of GD$_n$-flows and $n\times n$ KdV flows}\label{mu}

Let 
\begin{align*}
&S_0=\oplus_{i=1}^{n-1} \C \L^i,\\
&S_1= \oplus_{i=1}^{n-1} \C e_{ni}, 
\end{align*}
where $\L$ is defined by \eqref{ka}.
We have seen in section \ref{gd} that the map $\Psi:\calM(S_0)\to \calM(S_1)$ defined by Definition \ref{oa} is a bijection from the phase space of the $n\times n$ KdV to the phase space of the DS $A_{n-1}^{(1)}$ hierarchies. The phase space
 of the GD$_n$ hierarchy is
$$\frakP_n= \{P_\xi=\p_x^n-(\sum_{j=1}^{n-1} \xi_j \p_x^{j-1})\n \xi=\sum_{i=1}^{n-1} \xi_i e_{ni} \}.$$
It is clear that $\phi:\calM(S_1)\to \frakP_n$ defined by $\phi(L_\xi)= P_\xi$ is a bijection, where 
$L_\xi=\p_x-(J+\xi)$ and $ P_\xi= \p_x^n-\sum_{i=1}^{n-1} \xi_i \p_x^{i-1}$.  

  Drinfeld and Sokolov \cite{DriSok84} shows that the DS $A_{n-1}^{(1)}$ hierarchy is the GD$_n$ flows constructed from pseudo-differential operators under the bijection $\phi$. 
 Hence the $n\times n$ KdV flows correspond to the GD$_n$ flows under the bijection $\hat\Psi= \phi\circ \Psi$ from $\calM(S_0)$ to $\frakP_n$. 
 This section gives a different proof and an algorithm to construct the solution $u_f$ and its reduced frame of the $n\times n$ KdV hierarchy from a solution $P_{\xi(t)}$ of the GD$_n$ hierarchy such that $\hat\Psi(u_f)= P_\xi$.  We need this to show that the Virasoro action we obtain in section \ref{mr} is the same given in the physics literature and contained in the survey article \cite{vanM} by van Moerbeke.  

We first review the construction of GD$_n$ hierarchy from pseudo-differential operators (cf. \cite{Dic03}). 
For a pseudo-differential operator $\tau=\sum_{j\leq n_0} \tau_j \p_x^j$, we use the following notations:
\begin{align*}
\tau_+&= \sum_{j\geq 0} \tau_j \p_x^j, \quad \tau_-= \sum_{j<0}\tau_j \p_x^j,\\
\tau'&= \sum_{j\leq n_0} \tau_j' \p_x^j, \quad \tau_j'= \frac{\p\tau_j}{\p x},\\
\tau^{(i)}&= \sum_{j\leq n_0} (\p_x^i \tau_j) \p_x^j.
\end{align*}

The flows are defined as follows:
\beq
P_{t_j}= [(P^{j/n})_+, \, P], \quad j\geq 1, j\not\equiv n (\mod n).
\eeq
Here $P^{\frac{1}{n}}$ is the pseudo-differential operator whose $n$-th power is $P$.  Note that if $j=kn$ for some positive integer, then the flow vanishes.  

It is known (cf. \cite{vanM}) that if $P_{\xi(t)}=\p_x^n -\sum_{j=1}^{n-1} \xi_j(t) \p_x^{j-1}$ is a solution of the GD$_n$-hierarchy, then there is a pseudo-differential operator $\sigma(t)$ of the form 
\beq\label{dq}
\sigma(t)= \I + \sum_{j=1}^\infty \sigma_{-j}(t) \p_x^{-j}
\eeq
such that 
\begin{align} & P_{\xi(t)}= \sigma(t) \p_x^n \sigma^{-1}(t), \label{bhb}\\
&\frac{\p \sigma}{\p t_j}\, \sigma^{-1}= -(\sigma \p^j\sigma^{-1})_-. \label{bha}
\end{align}

Recall that the phase spaces of the $n\times n$ KdV hierarchy is $\calM(S_0)=\{L_u=\p_x -(J+ u)\n u\in C^\infty(\R,S_0)\}$ and the map $\hat\Psi: \calM(S_0)\to \frakP_n$ defined by
\beq\label{hu}
\hat\Phi(L_u)=P_\xi, \quad {\rm if\,\,}  \Psi(L_u)= L_\xi,
\eeq
  is a bijection.  By Proposition \ref{ba},  there exist unique smooth maps $\D:\R\to N_-$ and $\xi\in C^\infty(\R, S_1)$ satisfying 
$$\D (\p_x-(J+u))\D^{-1}= \p_x-(J+\xi).$$

\bthm\label{ej} The flows of the $n\times n$ KdV hierarchy correspond to the flows of the GD$_n$ hierarchy under the bijection $\hat\Psi:\calM(S_0)\to \frakP_n$ defined by $P(L_u)= P_\xi$, where $\xi$ is the GD variable for $u$.
\ethm

\bsub{\bf Outline of the proof of Theorem \ref{ej}}\par

\ss We identify $t_1$ with $x$.
 Given a solution $P(t)=\p_x^n-\sum_{i=1}^{n-1}\xi_i(t) \p_x^{i-1}$ of the GD$_n$ hierarchy, let $\sigma(t)$ be as in \eqref{dq} that satisfying \eqref{bhb} and \eqref{bha}, i.e., $P(t)= \sigma(t) \p_x^n \sigma(t)^{-1}$.  
 \ben
 \item[Step 1.] We use $\sigma(t)$ to construct a parallel frame $\ti F$ for the connection $(\p_x-(J+\xi))$, i.e., 
$\ti F_x \ti F^{-1}= J+\xi$, where $\xi= \sum_{i=1}^{n-1} \xi_i e_{ni}$. 
\item[Step 2.]  Set $Q(t)= \ti F(t) V(t)^{-1}$, where $V(t)=\exp(\sum_{j=1}^N t_j J^j)$ is the vacuum frame.  Factor $Q(t)= \eta_+(t) M(t)$ with $\eta_+(t)\in L_+$ and $M(t)\in L_-$, where $L_\pm$ is the splitting that gives the $n\times n$ KdV hierarchy.  Then we prove that 
\ben
\item[(i)] $\eta_+$ is in the subgroup $N_-$ of strictly lower triangular matrices and independent of $\l$,
\item[(ii)] $M_{t_j} M^{-1}= -(MJ^jM^{-1})_-$, for all $j\geq 0$ and $j\not\equiv 0$ ($\mod n$),
\item[(iii)] $(MJM^{-1})_+$ is of the form $J+u$ for some solution $u$ of the $n\times n$ KdV hierarchy.
\een
\item[Step 3.] We prove that $\xi$ is the GD variable for $u$. Hence $\hat\Psi$ maps the $n\times n$ KdV flows to the GD$_n$ flows.  
\een
\esub

The following Lemma proves Step 1. 

 \blem  \label{br}
Let $t_1=x$, $P_{\xi(t)}=\p_x^n-\sum_{i=1}^{n-1}\xi_i(t) \p_x^{i-1}$ a solution of the GD$_n$-hierarchy, and $\sigma(t)$  a pseudo-differential operator of the form \eqref{dq} satisfying \eqref{bhb} and \eqref{bha}.  Let $V_1$ denote the first row of $V(t)= \exp(\sum_{j=1}^N t_j J^j)$. Let $\ti F$ denote the matrix valued map whose first row $\ti F_1$ is $\sigma(t)(V_1)$ and the $(i+1)$-th row $\ti F_{i+1}= (\ti F_i)_x$ for $1\leq i\leq n-1$.  Then
 \ben
 \item $L_\xi \ti F=0$, where $L_\xi= \p_x-(J+\xi)$ and $\xi=\sum_{i=1}^{n-1}\xi_i e_{ni}$.
 \item $\ti F(t)= S(\sigma) V(t)$, where $S(\sigma)=(S_{ij}(\sigma))$ is defined by   by
 \beq\label{dp}
 S_{ij}(\sigma)=\bca 0, & {\rm if \ } i<j,\\
C_{i-1, j-1}\sigma^{(i-j)}, & {\rm if \ } i\geq j,\eca
 \eeq
 where $C_{i-1,j-1}=\frac{(i-1)!}{(j-1)!(i-j)!}$ is the binomial coefficient. 
 \een
 \elem

\begin{proof}\par

(1) We need to prove $\ti F_x= (J+\xi) \ti F$, i.e., 
\beq\label{mv}
\bca (\ti F_i)_x= \ti F_{i+1}, & 1\leq i\leq n-1,\\
(\ti F_n)_x= (\l+\xi_1) \ti F_1 + \xi_2 \ti F_2 + \cdots + \xi_{n-1} \ti F_{n-1}.
\eca
\eeq
The first equations of \eqref{mv} follows from the definition of $\ti F$. To prove the second equation of \eqref{mv} is equivalent to prove
\beq\label{mv3}
\p_x^n \ti F_1 = (\l+ \xi_1) \ti F_1 + \xi_2 \p_x \ti F_1 + \cdots + \xi_{n-1}\p_x^{n-2}\ti F_1.
\eeq
To prove this, we proceed as follows: 
Let $V_i$ denote the $i$-th row of $V$.  
Since $\p_x V= J V$, 
we have
\begin{equation}\label{bka}
 \bca \p_x V_i= V_{i+1}, &  1\leq i\leq n-1,\\
 \p_x V_n= \l V_1.\eca
 \end{equation}
So $\p_x^n V_1=\l V_1$.  But $P_\xi=\sigma \p_x^n \sigma^{-1}$ and $\ti F_1= \sigma(t) \ti V_1$ imply that 
$$P_\xi \ti F_1= P_\xi \sigma V_1=\l \ti F_1.$$ 
Hence $(P_\xi-\l)\ti F_1=0$.  
But
$$P_\xi-\l=\p_x^n -((\l + \xi_1) + \xi_2 \p_x + \cdots + \xi_{n-1}\p_x^{n-2}).$$
This proves \eqref{mv3} and $L_\xi\ti F=0$. 

(2) follows from a simple computation using 
$$\p_x \sigma= \sigma' + \sigma\p_x.$$
\end{proof}

\blem \label{baa} The operator $S$ defined in Lemma \ref{br} has the following properties:
\ben
\item[(i)] $S$ is linear,
\item[(ii)] $S(\sigma\tau)= S(\sigma)S(\tau)$,
\item[(iii)] $S(\sigma)= \sigma\I +\sum_{j=1}^{n-1} \sigma^{(j)}\,\frac{C^j}{j!}$, where $C=\sum_{k=1}^{n-1} ke_{k+1, k}$,
\item[(iv)] if $\sigma=\I +\sum_{i<0} \sigma_i \p_x^i$, then $S(\sigma)= \I +\sum_{i<0} S_i \p_x^i$, where 
\beq\label{er}
S_i= \sum_{j=0}^{n-1} \sigma_i^{(j)} \frac{C^j}{j!}.
\eeq
\een
\elem

\begin{proof}
It is clear that $S$ is linear.  The rest of the Proposition follows from simple and direct computations.
\end{proof}

Next we give a necessary condition for $g\in L_-(SL(n,\C))$ being in the negative subgroup $L_-$ of the $n\times n$ KdV hierarchy:

\blem\label{dta} Let $B:sl(n,\C)\to \calN_-$ be the linear operator \eqref{gg} defines the splitting $\calL_+, \calL_-=\calL_-^B$ of the $n\times n$ KdV hierarchy. If $g(\l)=\sum_{i\leq 0} g_i \l^i\in L_-$, then $g_0= \I+ B(g_{-1})$. 
\elem

\begin{proof}
We need to use the equivalent splitting of the $n\times n$ KdV hierarchy constructed in \cite{TerUhl11} to prove this Lemma:  Let $c_k= \prod_{i=1}^k \frac{1-\a^i}{1-\a}$, and 
\beq\label{hr}
\phi_n(z)= \sum_{k=0}^{n-1} c_k \L^k z^k,
\eeq
where $\a=\exp(2\pi i/n)$ and $\L$ is defined by \eqref{ka}.  It was proved in \cite{TerUhl11} that $\phi_n^{-1}(z)$ is a polynomial in $z$.  
Let $L^{\phi_n}$ denote the subgroup of $f\in L(SL(n,\C))$ satisfying the reality condition
\beq\label{fh}
\phi_n(z) f(z) \phi_n(z)^{-1}= \phi_n(\a z) f(\a z) \phi_n^{-1}(\a z), \quad \a= e^{2\pi i/n}.
\eeq
In other words, $\eta(z)=\sum_i \eta_i z^i\in \calL^{\phi_n}$ if and only if $\phi_n (z) \eta(z)\phi_n(z)^{-1}$ is a power series in $z^n$. 
Let $\Phi$ denote the following Lie algebra isomorphism: 
\beq\label{hq} 
\Phi:\calL(sl(n,\C))\to \calL^{\phi_n}, \quad \Phi(\eta)(z)= \phi_n(z)^{-1} \eta(z^n) \phi_n(z).
\eeq
Let $\calL_\pm^{\phi_n}= \calL_\pm(sl(n,\C))\cap \calL^{\phi_n}$, where $\calL_\pm(sl(n,\C))$ is the standard splitting of $\calL(sl(n,\C))$.  
The following results were proved in \cite{TerUhl11}:
\ben
\item[(i)] Let $\calL^{\phi_n}_\pm= \calL^{\phi_n}\cap \calL_\pm(sl(n,\C))$. Then $\Phi(\calL_+(sl(n,\C))= \calL^{\phi_n}_+$ and $\Phi(\calL_-)= \calL_-^{\phi_n}$, where $\calL_-$ is the negative group for the $n\times n$ KdV hierarchy. 
\item[(ii)]  If $\xi\in sl(n,\C)$, then 
\beq\label{hs}
\pi^{\phi_n}_+(\phi_n(z)^{-1} \xi z^{-n} \phi_n(z))= -\phi_n(z)^{-1}B(\xi) \phi_n(z),
\eeq
where $\pi_-^{\phi_n}$ is the projection of $\calL^{\phi_n}$ onto $\calL_-^{\phi_n}$ with respect to the splitting $\calL_\pm^{\phi_n}$. 
\item[(iii)] If $k= \I_n+\sum_{i<0} k_i z^i$ lies in $L_-^{\phi_n}$, then $k-\I_n\in \calL^{\phi_n}_-$.
\een

Given $g=\sum_{i\leq 0} g_i \l^i\in L_-^B$, first we claim that there exists $h(\l)= \I+\sum_{i\leq -2} h_i \l^i$ such that $gh= g_0+ g_{-1}\l^{-1}$ is in $L_-^B$ ($h_i$'s can be solved by comparing coefficients of $\l^i$).
It follows from (iii) of the results for the equivalent splitting of $n\times n$ KdV hierarchy in $z$-gauge given above that $\phi_n^{-1}(g_0+ g_{-1}z^{-n} -\I)\phi_n$ is in $\calL_-^{\phi_n}$, hence
$$\pi^{\phi_n}_+(\phi_n(z)^{-1}(g_0-\I_n+ g_{-1} z^{-n}) \phi_n(z))=0.$$
By \eqref{hs}, we have 
$$\pi^{\phi_n}_+(\phi_n(z)^{-1} g_{-1} z^{-n} \phi_n(z))= -\phi_n(z)^{-1} B(g_{-1}) \phi_n(z).$$
But $\phi_n(z)^{-1} (g_0-\I_n) \phi_n(z)$ is a polynomial in $z$. So we have
 $\phi_n^{-1} (g_0-\I_n -B(g_{-1}))\phi_n=0$.  This proves that $g_0-\I -B(g_{-1})=0$.
 \end{proof}

The next Lemma proves Step 2 (i).

\ss
\blem\label{ek} Let $\ti F(t)$ and $V(t)$ be as in Lemma \ref{br}, and $W= \ti F V^{-1}$. Then
\ben
\item[(i)] $W=\sum_{i\leq 0} q_i \l^i$ and $q_0\in N_-$,
\item[(ii)] factor $W=\eta_+ M$ with respect to the splitting of the $n\times n$ KdV, then $\eta_+$ is independent of $\l$ and lies in $N_-$. 
\een
\elem

\begin{proof}
It follows from Lemma \ref{baa} that 
$$S(\sigma) = \sum_{j\leq 0} S(\sigma_j\p_x^j) = \sum_{i=0}^{n-1} \frac{\sigma^{(i)}}{i!} C^i= \sum_{i=0}^{n-1} \sum_{j\leq 0} \frac{\sigma_j^{(i)}}{i!} C^i \p_x^j=\sum_{j\leq 0} S_j \p_x^j,$$
where 
$S_j = \sum_{i=0}^{n-1} \frac{\sigma_j^{(i)} }{i!} C^i$ and $C= \sum_{k=1}^{n-1} k e_{k+1,k}$ as given in Lemma \ref{baa}.
 
 Since $\p_x^j V= J^j V$, $$W=(S(\sigma)V)V^{-1}= (\sum_{j\leq 0} S_j\p_x^j V) V^{-1}= \sum_{j\leq 0} S_j J^j.$$ 
 
 To write $W$ as a power series in $\l$, we note that $J^{-n}=\l^{-1}$ and 
 $J^{-i}= (b^t)^i + b^{n-i} \l^{-1}$ for $1\leq i\leq n-1$.  Thus we have
 $$W= \I + \sum_{i=1}^{n-1} S_{-i} J^{-i} + \sum_{i\geq n} S_{-i} J^{-i}.$$
 The degree of the third term is $-1$.  So only the first and second term contribute to the constant term of $W$. 
 Since $S_{-i}\in \calN_-$ and the constant term of $J^{-i}$ is $(b^t)^i$, the constant term of $W$ in $\l$ expansion is in $N_-$. This proves (i).
 
 By Lemma \ref{dta}, $M$ is of the form 
 $$(\I+ B(p_{-1})) + p_{-1} \l^{-1} + p_{-2}\l^{-2} + \cdots.$$
 Equate the constant term in $W=\eta_+ M$ to see that $\eta_+$ is independent of $\l$ and $q_0= \eta_+(1+B(p_{-1}))$.  So $\eta_+= q_0(\I+ B(p_{-1}))^{-1}$, which is independent of $\l$ and is in $N_-$.
 \end{proof}

\blem \label{dr}
Let $\ti F= S(\sigma)V(t)$ as in Lemma \ref{br}, and $T$ a pseudo-differential operator.  Then    
\begin{equation}\label{bna}
((T_-\ti F)\ti F^{-1})_-=((T\ti F) \ti F^{-1})_-
\end{equation}
with respect to the splitting of the $n\times n$ KdV hierarchy.
\elem

\begin{proof}
Let $\ti E$ denote the solution of 
$$L_\xi \ti E=0, \quad \ti E(0,\l)=\I.$$
Then $\ti E(t, \cdot)\in L_+$. Hence $(T_+\ti E)\ti E^{-1}\in \ti \calL_+$.  Since $L_\xi\ti F=0$, there exists $f$ independent of all $t_j$'s such that $\ti F= \ti E f$.   So $(T_+\ti F)\ti F^{-1}= (T_+\ti E)\ti E^{-1}\in \calL_+$.   
But
$$
((T_-(\ti F))\ti F^{-1})_-= ((T(\ti F))\ti F^{-1}- (T_+(\ti F))\ti F^{-1})_-.
$$
We have shown that the second term is in $\calL_+$. Thus \eqref{bna} is true.
\end{proof}

Now we are ready to prove Step 2 (ii) and (iii).

\blem\label{bo} Let $\ti F(t)= S(\sigma)V(t)$ as in Lemma \ref{br}, and $W(t)=\ti F V(t)^{-1}$. Factor $W(t)= \eta_+(t) M(t)$ with $\eta_+(t)\in N_-$ and $M\in L_-$ as in Lemma \ref{ek}.  Then
\ben
\item[(i)] $M_{t_j} M^{-1}= -(MJ^jM^{-1})_-$,
\item[(ii)] $M(\p_x-J)M^{-1} =\p_x-(J+ u)$ for some solution $u$ of the $n\times n$ KdV hierarchy.
\een
\elem

\begin{proof}   
Note that  $S(\p_x^j) V= \p_x^j V= J^j V$. Use $M= \eta_+^{-1}W$ to get
\beq\label{el}
M_{t_j}M^{-1}= (\eta_+^{-1} W_{t_j} W^{-1}\eta_+)_- = (\eta_+^{-1} (W_{t_j}W^{-1})_- \eta_+)_-.
\eeq
Since $W= \ti F V^{-1}$ and $\ti F= S(\sigma)V$, it follows from Lemma \ref{baa} that we have
\begin{align*}
W_{t_j} W^{-1}&=\ti F_{t_j} \ti F^{-1}- W J^j W^{-1} = (S(\sigma)V)_{t_j} \ti F^{-1}- WJ^j W^{-1}\\
&= (S(\sigma_{t_j})V)\ti F^{-1}+ S(\sigma)V_{t_j} \ti F^{-1} - WJ^j W^{-1}, \quad {\rm by\, Lemma \, \ref{baa}\, (ii)}\\
& =(S(\sigma_{t_j}\sigma^{-1}) S(\sigma)  V + S(\sigma) J^j V)\ti F^{-1}  -W J^j W^{-1}\\
&= -(S((\sigma\p_x^j\sigma^{-1})_-)S(\sigma)V) \ti F^{-1} + (S(\sigma)J^j V)\ti F^{-1} -W J^j W^{-1}.
\end{align*}
By Lemma \ref{dr}, 
\begin{align*}
&((S((\sigma\p_x^j\sigma^{-1})_-)S(\sigma) V) \ti F^{-1})_-= -((S(\sigma\p_x^j \sigma^{-1})S(\sigma)V)\ti F^{-1})_- \\
&= -((S(\sigma\p_x^j)V)\ti F^{-1})_-=  -(S(\sigma)S(\p_x^j) V)\ti F^{-1})_- = -((S(\sigma) J^j V)\ti F^{-1})_-.
\end{align*}
So we have
$$W_{t_j} W^{-1}\equiv  - W J^j W^{-1}\quad (\mod \calL_+).$$
Here we use the notation $\xi\equiv \eta$ $ (\mod \calL_+)$ if $\xi-\eta\in \calL_+$. 

By \eqref{el}, we have 
\begin{align*}
&M_{t_j}M^{-1} = (\eta_+^{-1}W_{t_j} W^{-1} \eta_+)_-= (\eta_+^{-1} (W_{t_j}W^{-1})_-\eta_+)_-\\
& = -(\eta_+^{-1} (W J^jW^{-1})_- \eta_+)_-= -(\eta_+^{-1} W J^j W^{-1}\eta_+)_- = -(M J^j  M^{-1})_-.
\end{align*}
This proves (i).  

Statement (ii) follows from (i) and Theorem \ref{ht}.
\end{proof}

\ss
\bsub{\it Proof of Theorem \ref{ej}}\par

 Let $P_{\xi(t)}$ be a solution of the GD$_n$-hierarchy, and $ \sigma(t),  V(t)$ and $\ti F(t)=S(\sigma(t))V(t)$ as in Lemma \ref{br}, and $W(t)= \ti F(t) V(t)^{-1}$.  Factor $W(t)= \eta_+(t) M(t)$ and let $u(t)$ as in Lemma \ref{bo}.  By Lemma \ref{bo}, $u(t)$ is a solution of the $n\times n$ KdV hierarchy.  
Let $L_u=\p_x-(J+u)$, and $L_\xi= \p_x-(J+\xi(t))$.  It remains to prove that $\Psi(L_u)=L_\xi$.  To prove this, we compute 
\begin{align*}
&(MV)_x(MV)^{-1}= M_x M^{-1}+ MV_xV^{-1}V^{-1} = M_xM^{-1} + MJM^{-1}\\
& \quad = -(MJM^{-1})_-+ MJM^{-1}= (MJM^{-1})_+ = J+u.
\end{align*}
By Lemma \ref{br}, we have $\ti F= S(\sigma)V$ and $\ti D\ti F=\ti F_x-(J+\xi)\ti F=0$. So $\ti F_x\ti F^{-1}= (J+\xi)$.  But $\ti F= W V= \eta_+ MV$.  So we have
$$\ti F_x\ti F^{-1}= (\eta_+)_x\eta_+^{-1} + \eta_+ (MV)_x(MV)^{-1} \eta_+^{-1}= (\eta_+)_x\eta_+^{-1}+ \eta_+(J+u)\eta_+^{-1}.$$
This implies that $\eta_+(\p_x-(J+u))\eta_+^{-1}= \p_x -(J+\xi)$.  By Lemma \ref{ek}, $\eta_+$ is in $N_-$ and independent of $\l$. So it follows from Corollary \ref{ob} that $\eta_+= \D$ and $\Psi(L+u)=L_\xi$. This completes the proof of Theorem.  \qed
\esub

The above proof in fact gives an algorithm to compute the solution of the $n\times n$ KdV solution that corresponds to the solution $\xi$ of the GD$_n$.

\bthm\label{gia}
Let $x=t_1$, $P_{\xi(t)}= \p_x^n-\sum_{i=1}^{n-1} \xi_i(t) \p_x^{i-1}$  a solution of the GD$_n$ hierarchy, and $\xi(t)= \sum_{i=1}^{n-1}\xi_i(t) e_{ni}$.  Let $\sigma(t)=\I+ \sum \sigma_j \p_x^{-j}$ such that $P_\xi= \sigma \p_x^n \sigma^{-1}$ and $\sigma_{t_j} \sigma^{-1}= -(\sigma \p_x^j \sigma^{-1})_-$.  Let $V(t)=\exp(\sum_{i=1}^N t_i J^i)$, and $S(\sigma)$ be defined by \eqref{dp}. Set $W=( S(\sigma)(V))V^{-1}$.  Factor $W= \D M$ with $\D\in L_+$ and $M\in L_-$. Then 
\ben
\item $\D_\l=0$ and $\D\in N_-$,
\item $u=MJM^{-1}-J$ is a solution of the $n\times n$ KdV hierarchy,
\item let $f=M(0)$, then $u=u_f$ and $M$ is the reduced frame,
\item $\hat\Psi(L_u)=P_\xi$, where $\hat\Psi$ is the bijection from $\calM(S_0)$ to $\frakP_n$ defined by \eqref{hu}.
\een
\ethm

\bs
\section{The Virasoro actions agree}\label{eo}

We gave an action of $\calV_+$ on reduced frames of the $n\times n$ KdV hierarchy in section \ref{mr}.  There is also a well-known  action of $\calV_+$ on the solutions of the GD$_n$-flows.  In this section, we prove that these two actions are the same via the isomorphism $\hat\Psi$ between the phase spaces of the $n\times n$ KdV and the GD$_n$ hierarchies given in section \ref{mu}.  

The well-known $\cv_+$-action  (cf. \cite{vanM}) on 
$$\frakS=\{\sigma=\I +\sum_{i<0} \sigma_i \p_x^i\n \sigma_i\in C^\infty(\R, \C)\}$$
is defined by
$$Y_\ell(\sigma)\sigma^{-1}= \left(\sigma\circ \left(x +\sum_{j=2}^N jt_j \p^{j-1}\right)\circ \p^{\ell+1}\circ \sigma^{-1}\right)_-, \quad \ell\geq -1.$$
Let 
$$\frakS_n=\{\sigma\in \frakS\n (\sigma\p_x^n\sigma^{-1})_-=0\}.$$ 
Then $\sigma\in \frakS_n$ if and only if $\sigma\p_x^n\sigma^{-1}\in \frakP_n$, the phase space of  the GD$_n$ hierarchy.  
It can be checked easily that the vector field 
$$\eta_\ell:= Y_{n\ell}$$ is tangent to $\frakS_n$ for $\ell\geq 0$. Hence the following vector fields
\beq\label{fw}
(\eta_\ell \sigma)\sigma^{-1}= \frac{1}{n}\left(\sigma\circ \left(x +\sum_{j=2}^N jt_j \p^{j-1}\right)\circ \p^{n\ell+1}\circ \sigma^{-1}\right)_-, \quad \ell\geq 0.
\eeq 
on $\frakS_n$ satisfy the Virasoro condition $[\eta_\ell, \eta_k]= (k_-\ell) \eta_{k+\ell}$ for $k, \ell \geq 0$. 

Next we compute the variation on the reduced frame $M$ induced from a variation of pseudo-differential operators:

\bprop\label{fx} Let $P_{\xi(t)}, \sigma(t), S(\sigma), M, \D$ be as in Theorem \ref{gia}.  If the variation $(\d\sigma)\sigma^{-1}= (\tau\sigma^{-1})_-$ for some pseudo-differential operator $\tau$, then the corresponding variation on $M$ is given by
\beq\label{fy}
(\d M)M^{-1}= ( \D^{-1}(S(\tau)V)V^{-1} M^{-1})_-.
\eeq
\eprop

\begin{proof}
Take variation of $M=\D^{-1} W$ to get
$$(\d M)M^{-1} = -\D^{-1}\d\D +\D^{-1} \d W W^{-1}\D.$$
Since $\D\in L_+$, $\D^{-1}\d \D\in \calL_+$. So we have
$$(\d M)M^{-1}= (\D^{-1} (\d W )W^{-1})\D)_-.$$
Use $W= (S(\sigma)V)V^{-1}$ and Lemma \ref{baa} to compute
\begin{align*}
\d W W^{-1}&= (S(\d \sigma)V) V^{-1}W^{-1}= (S((\tau\sigma^{-1})_-\sigma)V)\ti F^{-1}\\
& = S((\tau\sigma^{-1})_-)S(\sigma)V) V^{-1}=(S((\tau \sigma^{-1})_-\ti F)\ti F^{-1}).
\end{align*}
By Lemma \ref{dr}, we have 
\begin{align*}
((\d W) W^{-1})_-&= (S(\tau\sigma^{-1})\ti F) \ti F^{-1})_-= ((S(\tau\sigma^{-1})S(\sigma)V)\ti F^{-1}\\
&= ((S(\tau)V)\ti F^{-1})_-.
\end{align*}
Therefore we get
\begin{align*}
(\d M)M^{-1}&= (\D^{-1}(\d W) W^{-1}\D)_-= (\D^{-1} ((\d W) W^{-1})_- \D)_-\\
&=(\D^{-1} (S(\tau)V)\ti F^{-1})_- \D)_- = (\D^{-1}(S(\tau) V)\ti F^{-1})\D)_-\\
&= (\D^{-1} (S(\tau) V) V^{-1}M^{-1})_-
\end{align*}
\end{proof}

\bthm\label{fxa} With the same notation as in Proposition \ref{fx},
if $\eta_\ell\sigma$ is the variation given by \eqref{fw}, then the corresponding variation on $M$ is 
\begin{align*}
& (\eta_\ell M)M^{-1}=\frac{1}{n}(Z_1+ Z_2+ Z_3 + Z_4)_-, \quad {\rm where\,\,}\\
&Z_1= xMJ^{n\ell+1}M^{-1},\quad 
Z_2=\sum_{j=2}^N jt_j MJ^{n\ell+ j} M^{-1},\\
&Z_3=-\sum_{j>0} j \D^{-1}S_{-j}J^{n\ell -j} M^{-1},\quad 
Z_4= \D^{-1}C\D M J^{n\ell+1}M^{-1},
\end{align*}
$C=\sum_{k=1}^n ke_{k+1,k}$, and $S_{-j}$ is defined by \eqref{er}. 
\ethm


\begin{proof}
We assume $(\d_\ell \sigma)\sigma^{-1}= \frac{1}{n}(\sigma\tau)_-$, where
$$\tau= \sigma\circ \left(x+ \sum_{j=2}^N jt_j\p^{j-1} \right) \p^{n\ell+1}.$$
Write $\sigma= \sum_{j\leq 0} \sigma_j \p^j$ with $\sigma_0=1$.
Then 
$$\sigma \circ x= x\sigma + \sum_{j<0} j \sigma_j \p^{j-1}.$$
So we have
\begin{align*}
\tau&= \sigma\circ x \circ \p^{n\ell +1} +\sum_{j=2}^N j t_j \sigma\circ \p^{n\ell+ j}\\
&= x\sigma\circ \p^{n\ell +1} - \sum_{j>0} j\sigma_{-j} \p^{n\ell-j} + \sum_{j=2}^N jt_j \sigma\circ \p^{n\ell+ j}.
\end{align*}
By Lemma \ref{baa} (iii), we have
\beq\label{ga}
S(\p^k)= \p^k \I, \quad S(x\I) = x\I + C, \quad S(\sigma_{-j} \p^m)= S_{-j} \p^m,
\eeq
where $C= \sum_{k=1}^{n-1} k e_{k+1, k}$ and $S_{-j}$ is defined by \eqref{er}.  

Use $\p_x^k V= VJ^k$ to get 
\beq\label{gca}
S(\tau)V= (x+C)S(\sigma)VJ^{n\ell+1} -\sum_{j=1}^\infty j S_{-j}J^{n\ell-j} V +\sum_{j=2}^N jt_j S(\sigma)V  J^{n\ell+j}.
\eeq
The Theorem follows by substituting \eqref{gca} to \eqref{fy}.
\end{proof}

\bthm\label{gi} The Virasoro actions on the $n\times n$ KdV flows and the GD$_n$ flows agree under the bijection $\hat \Psi$.
\ethm

\begin{proof} By Theorem \ref{fb}, the induced $\cv_+$-action on reduced frame $M$ of the $n\times n$ KdV hierarchy is 
\beq\label{ep}
(\d_\ell M)M^{-1}= -(\l^\ell (\G M)M^{-1} + \frac{1}{n}\l^\ell M\sum_{i\geq 1} i t_i MJ^i M^{-1})_-, \quad \ell\geq -1.
\eeq
Recall that the Virasoro vector fields on the reduced frames $M$ induced from the Virasoro vector fields $\eta_\ell$ on $\frakS_n$'s is 
$$(\eta_\ell M)M^{-1}= \frac{1}{n}(Z_1+ Z_2+ Z_3+ Z_4)_-$$
as given in Theorem \ref{fxa}.  We want to prove that $\eta_\ell= \d_\ell$.  
It is clear that the second term of \eqref{ep} is equal to $(Z_1+Z_2)_-$. 

To compute the first term of \eqref{ep}, we recall that $\hat \G f= \l f_\l + [f, \Xi]$ is the derivation defined by \eqref{dm3}. Let $W, \D$ be as in Theorem \ref{gia}. 
By Definition of $\hat \G$, we have 
$$(\hat \G W) W^{-1} = (\G W)W^{-1} - \Xi$$

Since $M= \D^{-1}W$ and $\D_\l=0$, a direct computation implies that
$$(\G M) M^{-1}= \D^{-1} (\G W) W^{-1}\D.$$ 
Hence we have
\begin{align*} (\l^\ell(\G M)M^{-1})_- &=(\l^\ell \D^{-1} (\G W)W^{-1} \D)_-\\
&= (\l^\ell \D^{-1} (\hat \G W)W^{-1}\D -\l^\ell \D^{-1} \Xi\D)_- \\
&= (\l^\ell \D^{-1} (\hat \G W)W^{-1}\D)_-.
\end{align*}
The last equality holds because $\ell >0$ implies that $\l^\ell \D^{-1} \Xi \D\in \calL_+$.  

By Lemma \ref{ek}, 
$W= \sum_{j\geq 0} S_{-j} J^{-j}$, and $S_{-j}$ is given by \eqref{er}, i.e., 
$$S_{-j}= \sum_{i=0}^{n-1} \frac{\sigma_{-j}^{(i)}}{i!} C^i,$$
where $C= \sum_{k=1}^{n-1} k e_{k+1, k}$.   

A simple computation shows that
$\hat\G C= -\frac{1}{n}C$ and $\hat\G C^k= -\frac{k}{n} C^k$.  Therefore
\beq\label{gj}
\hat\G S_{-j}=-\frac{C}{n} (S_{-j})_x.
\eeq

Use \eqref{gj} and $\hat\G J^k=\frac{k}{n} J^k$ (Lemma \ref{cv}) to get 
$$\hat\G W=\frac{1}{n} \sum_{j>0} -C(S_{-j})_x J^{-j} - jS_{-j}J^{-j}.$$
Therefore we have proved
$((\G M)M^{-1}\l^{\ell})_-= \frac{1}{n} (Y_3+ Y_4)_-$, where
\begin{align*}
Y_3&= - \l^{\ell}\sum_{j>0} j \D^{-1}S_{-j}J^{-j} M^{-1},\\
Y_4&= - \l^{\ell}\D^{-1} C (S_{-j})_x J^{-j} M^{-1}. 
\end{align*}
It is clear that $(Y_3)_-= (Z_3)_-$.  It remains to prove $(Y_4)_-= (Z_4)_-$.  Since
$$M=\D^{-1}W= \D^{-1}\sum_{j\geq 0}S_{-j}J^{-j},$$
$$M_x M^{-1} = -\D^{-1}\D_x + \D^{-1}\sum_{j>0} (S_{-j})_x J^{-j} M^{-1}.$$
This implies that
$$\sum_{j>0} (S_{-j})_x J^{-j}M^{-1} = \D M_x M^{-1} +\D_x.$$
Hence we have
$$
 (Y_4)_-=- (\l^{\ell} \D^{-1} C \D M_x M^{-1}+\D^{-1}\D_x)_-= -(\l^{\ell} \D^{-1} C \D M_x M^{-1})_-.$$
 But $M_{t_1} M^{-1}= -\pi_-(MJM)$ (recall that we identify $t_1= x$). So we have
  $$(Y_4)_-= (\l^{\ell}\D^{-1} C \D (MJM^{-1})_-)_-=\pi_-(\l^{\ell}\D^{-1} C\D MJM^{-1})_-= (Y_4)_-.$$
This completes the proof. \end{proof}

\bs

\end{document}